\documentclass[twocolumn]{aastex631}

\usepackage{array}
\usepackage{longtable}
\usepackage{amsmath}

\begin{document}

\title{Counterculture Stars: Slow and Retrograde Stars with Low-Alpha Disk Abundances}

\correspondingauthor{Carrie Filion}\email{cfilion@flatironinstitute.org}
\author[0000-0001-5522-5029]{Carrie Filion}
\affiliation{Center for Computational Astrophysics, Flatiron Institute, 162 Fifth Avenue, New York, NY 10010, USA}

\author[0000-0003-1517-3935]{Michael S. Petersen}
\affiliation{Institute for Astronomy, University of Edinburgh, Royal Observatory Edinburgh, Blackford Hill, Edinburgh EH93HJ, UK}

\author[0000-0003-1856-2151]{Danny Horta}
\affiliation{Institute for Astronomy, University of Edinburgh, Royal Observatory Edinburgh, Blackford Hill, Edinburgh EH93HJ, UK}
\affiliation{Center for Computational Astrophysics, Flatiron Institute, 162 Fifth Avenue, New York, NY 10010, USA}

\author[0000-0003-2594-8052]{Kathryne J. Daniel}
\affiliation{University of Arizona Department of Astronomy and Steward Observatory, 933 N Cherry Ave., Tucson, AZ 85719, USA}

\author[0000-0001-7297-8508]{Madeline Lucey}
\affiliation{
Department of Physics \& Astronomy, University of Pennsylvania, 209 S 33rd St., Philadelphia, PA 19104, USA }

\author[0000-0003-0872-7098]{Adrian~M.~Price-Whelan}
\affiliation{Center for Computational Astrophysics, Flatiron Institute, 162 Fifth Avenue, New York, NY 10010, USA}



\begin{abstract}

The Milky Way is home to a thin disk that can be defined via kinematics and/or elemental abundances. The elemental abundance-defined thin disk, also called the low-alpha disk, is generally thought to be comprised of stars on planar, circular orbits that approximate the circular velocity curve. While this is an apt description for the majority of stars with thin-disk-like abundances, there are a number of interesting exceptions. In this analysis, we identify and investigate $\sim 70$ stars with thin-disk-like abundances and very slow or retrograde Galactocentric azimuthal velocities. These stars could be kinematical outliers of the thin disk or elemental abundance outliers of the halo. Focusing first on the former, we introduce a number of mechanisms that could alter a thin disk orbit and cause the azimuthal velocity to become slow or retrograde. We then determine signatures for each mechanism and assess whether that mechanism is unlikely, plausible, or consistent given each star's reported properties. We find that at least one mechanism is plausible for each star, and the mechanism with the highest number of consistent candidate stars is dynamical ejection from stellar clusters. We next discuss scenarios that could produce halo stars with thin disk abundances, and again identify stars that could be connected to these mechanisms. With this sample we investigate rare processes, such as binary disruption by the central supermassive black hole, while also providing a unique perspective into the chemo-dynamics and structural components of the Milky Way.

\end{abstract}


\section{Introduction} \label{sec:intro}

The Milky Way hosts a number of structural components with distinct but overlapping properties, such as the thin and thick disks. As the names imply, the thin disk has a lower scale height and is kinematically cooler than the thick disk \citep[e.g.,][]{Gilmore1989, Wyse1995, Bovy2012, Haywood2013, Rix2013, bland_hawthorn_2016}. These disks can be characterized kinematically, geometrically and/or chemically, though each characterization does not necessarily define the same populations. Elemental abundance-based definitions tend to yield more cleanly separated kinematics than vice-versa \citep[e.g.,][]{Nidever2014, Hayden2015, Mackereth2017,Carrillo2024}, and we here adopt this definition. Stars in the elemental abundance-based definition of the thin disk have low or no enhancement in the alpha elements (i.e. elements predominantly produced in type II supernovae) relative to iron, and the thin disk is thus commonly referred to as the \lq alpha-poor' or \lq low-alpha' disk \citep[e.g.,][]{Weinberg2019}. In addition to the stellar disks, there is a stellar halo that is constructed largely of old and metal-poor stars on generally non-planar orbits \citep[e.g.,][]{Eggen1962, Gilmore1998, Chiba2000, Carollo2010}, proposed to be at least partly comprised of accreted satellite galaxies \citep[e.g.,][]{Wyse2006,belokurov_2018, helmi_2018, Myeong2019,mackereth_2019,naidu_2020, horta_2021}.

Kinematical selections of thin disk stars often require that stars be on approximately circular orbits close to the Galactic plane, such that the velocity in the Galactocentric, cylindrical radial and vertical velocities are near zero ($V_R \sim V_z \sim 0 $ km~s$^{-1}$), and the Galactocentric azimuthal velocity is close to the circular velocity curve (i.e. $|V_\phi|$ $\sim V_c \approx 230$ km~s$^{-1}$ in the Solar neighborhood; \citealp{eilers_2019}). The vast majority of stars with low-alpha (thin) disk abundances have orbits that fit these requirements \citep[][]{mackereth_2019}. In this analysis, we investigate those stars that have thin-disk-like abundances but \textit{do not} have thin-disk-like kinematics. Specifically, we are interested in stars with slow or retrograde azimuthal velocities.

Here we define \lq retrograde' to be an instantaneous velocity in the azimuthal ($\phi$) direction that is opposite in sign from the mean rotation of the disk. We adopt this observationally-motivated definition as it requires the fewest number of assumptions: from just observed positions, velocities, and a definition of Galactocentric coordinate frame, we can determine that a star has an azimuthal velocity ($V_\phi$) that is opposite in sign to the disk rotation. Integrating an orbit to determine whether it is retrograde with respect to (e.g.) the bar requires knowledge of the Galactic potential and basic properties of the structure in the Milky Way, such as the length and speed of the bar, are still hotly debated \citep[for example,][]{Portail2017,Sanders2019,Leung2023,lucey_2023,Zhang2024}. To avoid artificially truncating the sample of extreme $V_\phi$ outliers, we extend the selection to include slow prograde stars ($\sim 6 \sigma$ slower than the median $V_\phi$ velocity).

Slow and retrograde stars with low-alpha, thin-disk-like elemental abundances represent either the extreme kinematical tail of the disk or the extreme abundance tail of the stellar halo. They could have come to their present chemodynamical state in a variety of ways, and indeed it is likely that multiple mechanisms have contributed to this population of stars. If a given star was born in the thin disk, its orbit must have been altered to slow or reverse the $V_\phi$ velocity, and we outline the possible mechanisms for such a change in this analysis. Alternatively, it is possible that a given star could have been born with thin-disk-like abundances but on its present halo-like orbit, and we discuss this possibility as an alternate scenario for some of the stars in the sample. For the former scenario, we consider dynamical cluster ejection, binary supernovae ejection, supermassive black hole ejection (also called Hills mechanism, \citealt{hills_1988}), early clumpy disk formation, interactions with the Galactic bar, and satellite passage as possible mechanisms for altering thin disk stellar orbits.

This manuscript is organized as follows: in Section \ref{sec:selection}, we outline the sample selection, as well as the properties of the sample with respect to the stellar populations of the Milky Way. In Section \ref{sec:mechanisms}, we introduce the possible mechanisms for producing slow and retrograde stars with thin-disk-like abundances, their possible observational signatures, and which (if any) stars in the sample could be linked to each mechanism. We discuss other perspectives in Section \ref{sec:discussion}, and conclude in Section \ref{sec:conclusion}.

\section{Observational Data}\label{sec:selection}

We first identify stars that 1) have abundances consistent with that of the low-alpha, thin disk 2) have $V_\phi$ velocities that are slowly prograde or retrograde.  We perform this search in the set of stars with homogeneously-derived, individual elemental abundances from APOGEE DR17 \citep{majewski, abdurrouf_2022} and full 6-D phase space information from the combination of APOGEE line-of-sight velocities \citep{nidever_2015} and \textit{Gaia} astrometry \citep{gaia_2016,gaia_2021, gaia_astrometry}. We use the abundance values from the ASPCAP pipeline \citep{garcia_perez_2016}, which uses the FERRE code\footnote{available on github: \url{https://github.com/callendeprieto/ferre}} \citep{allende_prieto_2006} and linelists presented in \cite{shetrone_2015} and \cite{smith_2021}. The spectrographs and Northern and Southern hemisphere telescopes are detailed in \cite{wilson_2019}, \cite{gunn_2006}, and \cite{bowen_1973}, respectively, and the targeting is described in \cite{zasowski_2013}, \cite{zasowski_2017}, \cite{beaton_2021}, and \cite{santana_2021}. We use the \textit{Gaia} information that is included in the ASPCAP catalog, based on a positional cross-match.

We enforce a number of quality cuts on both the spectroscopic and astrometric parameters. For the spectroscopic parameters from APOGEE, we first require that the stars be relatively warm red giants with $4000 < \rm{T}_{eff} < 5500$~K, $0 < \rm{log(g)} < 3$, log(g) error $< 0.2$, and signal to noise $>50$. We remove entries that are not the highest quality, primary science targets by requiring that \texttt{EXTRATARG} = 0, and exclude stars with problematic spectra (those marked as bad, stars with very bright neighbors, etc.) by requiring that \texttt{ASPCAPFLAG} bits are not set to 23, and that the \texttt{STARFLAG} bits are not set to 0, 1, 3, 16, 17, 19, 21, or 22. We remove likely members of clusters and dwarf galaxies by requiring \texttt{MEMBERFLAG} $ = 0$. Finally, we require that the flags for each of the abundances that we employ in this work are set to zero. We consider stars that pass these cuts to have quality spectra and abundance information. 

For the astrometric parameters, we require parallax $>0$, and that the errors be $<10\%$ in the parallax and both components of the proper motion. Stars that pass both the spectroscopic and astrometric cuts are considered to have quality abundance and kinematic information. This quality sample is not complete and is biased towards more local stars, but is suitable for the present purpose. The final sample has line-of-sight velocity uncertainties of less than ten percent.

\begin{figure*}   \includegraphics[width=\textwidth]{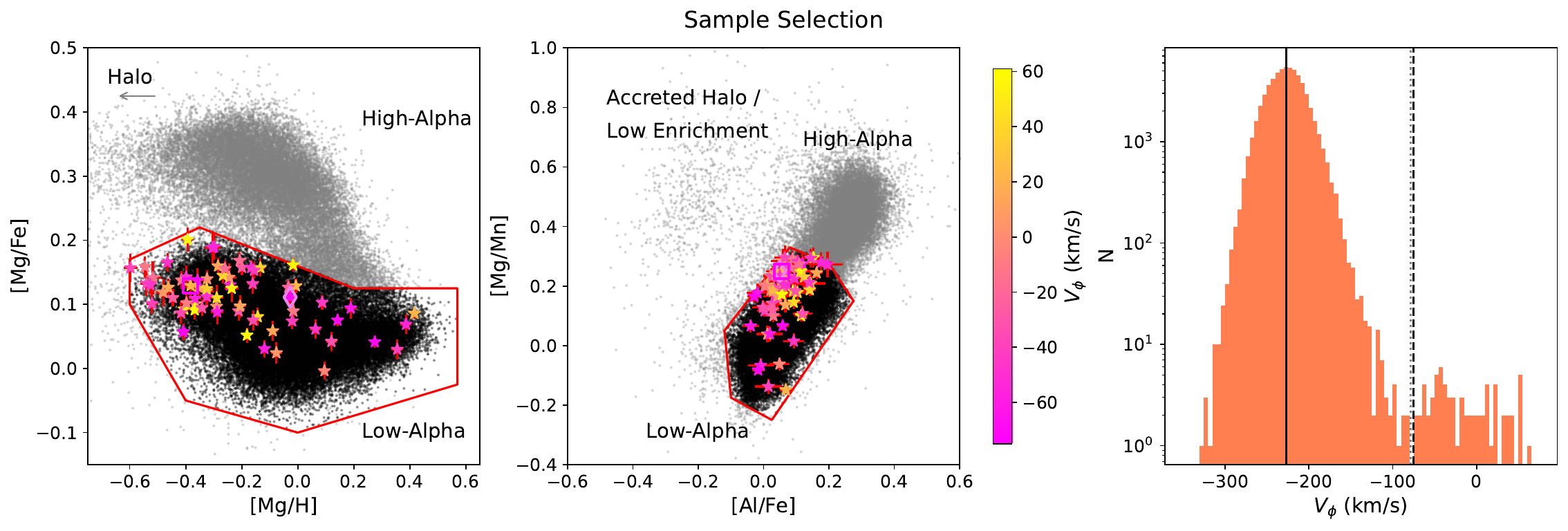}
    \caption{The [Mg/H] vs [Mg/Fe] (left) and [Al/Fe] vs [Mg/Mn] (middle) abundances for all quality stars (grey), stars with thin disk abundances (black), and the sample of slow and retrograde stars with thin disk abundances (larger star-shaped points color-coded by azimuthal velocity $V_\phi$), alongside the $V_\phi$ velocity distribution for all quality thin disk stars (right). The solid black line in the left panel denotes the median $V_\phi$ velocity of the thin disk stars, the dotted grey line is the median plus six standard deviations, and the dashed black line is -75 km~s$^{-1}$. In the left and middle plot, APOGEE abundance errors for the slow and retrograde stars are shown in red, where the [Mn/Mg] and [Mg/H] error bars are the quoted APOGEE errors for the constituent terms added in quadrature. The selection criteria for the thin-disk cut are shown as red polygons. The thin disk sample must have abundances within both of these polygons, while the slow and retrograde sample must qualify these abundance criteria as well as kinematic criteria. The open magenta square identifies a likely binary star, while the open violet diamond identifies a fast rotator. These square and diamond identifiers are the same throughout the figures.}
    \label{fig:alpha_cut}
\end{figure*}

We define stars with ``thin disk abundances'' to be those within both of the polygons that we define (by eye) in the left and middle panels of Figure \ref{fig:alpha_cut}. The polygon in the left panel of Figure \ref{fig:alpha_cut} identifies the typical locus of low-alpha, thin disk stars in the [Mg/H]--[Mg/Fe] plane, where we adopt the convention of (e.g.) \cite{weinberg_2019} and employ [Mg/H] as the x-axis as opposed to [Fe/H]. The polygon in the middle panel in Figure \ref{fig:alpha_cut} indicates the typical location of low-alpha, thin disk stars in the [Al/Fe]--[Mg/Mn] plane\footnote{The coordinates of the polygons are: [ [-.6, .17], [-.35, .22], [0.2, .125], [.57, .125], [.57, -.025], [0, -.1], [-.4, -.05], [-.6, .1],[-.6, .17]] for the [Mg/H]--[Mg/Fe] plane and [[-.12, .05], [.08,.33], [.2, .275], [.275, .15], [0.025, -.25], [-.1, -.175],[-.12, .05]] for the [Al/Fe]--[Mg/Mn] plane}. Typically, stars in the upper left quadrant are considered to be from accreted (or chemically unevolved, e.g. \citealt{horta_2021}) populations, while stars in the blob to the upper right of the polygon are thick disk stars (see e.g. \citealt{hawkins_2015} and labels in Figure \ref{fig:alpha_cut}). We henceforth assume that the \lq low-alpha' disk and stars with \lq thin-disk-like abundances' are synonymous. We prefer the term \lq thin disk abundances' over \lq low-alpha' as it is a more accurate descriptor: we use two elemental abundance planes to make our selection and low-alpha stars are not necessarily always within [Al/Fe]--[Mg/Mn] polygon. The additional selection in [Al/Fe]--[Mg/Mn] distinguishes our sample from earlier analyses interested in (e.g.) metal-rich halo stars or generic retrograde samples \citep[for example, the analyses presented in][]{bonaca_2017, kordopatis_2020, ceccarelli_2024}.

We also use the stellar age estimates provided in the \texttt{astroNN} \citep{astro_nn}\footnote{GitHub page available at \url{https://github.com/henrysky/astroNN}} value added catalog\footnote{see \url{https://www.sdss4.org/dr17/data_access/value-added-catalogs/?vac_id=the-astronn-catalog-of-abundances,-distances,-and-ages-for-apogee-dr17-stars}}. \texttt{astroNN} is a neural network that has been trained on and applied to the APOGEE data to provide estimates for abundances, ages, stellar parameters, etc. In this analysis, we use the recommended \texttt{age\_lowess\_correct} ages (see e.g. \citealt{mackereth_2019}). The average \texttt{age\_total\_error} for the final sample is $\sim 1.7$ Gyr, while the maximum is $3.4$ Gyr. We use these age estimates mainly to group stars into old versus young, and stress that the accuracy of any given individual age estimate cannot be guaranteed. As \texttt{astroNN} uses spectra, atypical stellar abundances, in particular, may impact a given age estimate.

We use Astropy \citep{astropy:2013,astropy:2018,astropy:2022} and Gala \citep{gala} to transform the astrometric data into Galactocentric coordinates using the \textit{Gaia} proper motions, APOGEE line-of-sight velocities (\texttt{VHELIO\_AVG}), and \cite{bailer_jones} distance estimates (\texttt{GAIAEDR3\_R\_MED\_PHOTOGEO}) provided in the ASPCAP catalog. Throughout this analysis we refer to the Galactocentric Cartesian x, y, z velocities as U, V, W and the Galactocentric cylindrical R, $\phi$, z velocities as $V_R$, $V_\phi$, $V_z$. We use values for the Solar position and velocity given in \citet[see references therein for derivations etc]{hunt_2022}, which positions the Sun at $R = 8.275$~kpc and $z = 20.8$~pc, with a velocity with respect to the Galactic center of ($U, V, W$) = (8.4, 251.8, 8.4) km~s$^{-1}$. The rotation of the disk is in the negative $V_\phi$ direction, such that retrograde stars have \textit{positive} $V_\phi$ values.

We next identify kinematic outliers, accounting for astrometric errors. We first select all stars with $V_\phi$ values that are more than $3$ standard deviations slower than the median $V_\phi$ of the thin disk, and then compute 100 realizations of the $V_\phi$ velocity of each of these stars. In each realization, we sample new astrometric information from Gaussian distributions centered on the reported values with standard deviation equal to the reported errors. In recomputing the $V_\phi$ values, we ignore covariances, keep RA, Dec fixed, and use \texttt{GAIAEDR3\_R\_HI\_PHOTOGEO} - \texttt{GAIAEDR3\_R\_LO\_PHOTOGEO} for the distance error. We consider the median of the 100 realizations to be the $V_\phi$ measurement, and the final sample of kinematical outliers are those with $V_\phi > -75$ km~s$^{-1}$. This threshold is approximately six standard deviations from the median $V_\phi$ velocity of the thin disk sample and is an easy, whole number limit to carry forward. 

There are sixty-nine stars in the final sample of slow and retrograde stars with thin-disk-like abundances. We will compare this sample to two subsets of the APOGEE data: the first is all stars with quality abundances and astrometry (\lq quality stars'), the second is the subset of those quality stars that have thin disk abundances (i.e. within the polygons in the left and middle panels of Figure \ref{fig:alpha_cut}, hereafter the \lq thin disk' sample). The sample of slow and retrograde stars with thin disk abundances represents $\sim 0.11\%$ of the thin disk sample as defined above. Alternatively, this final sample represents $\sim 4\%$ of all of the quality stars in APOGEE with $V_\phi$ velocities above -75 km~s$^{-1}$. Only one star in the final sample has a scatter in its line-of-sight velocity measurements indicative of possible binarity, and only one star has a reported (high) rotation. These stars are indicated with a magenta square and violet diamond, respectively, in all of the plots where they are present.

\subsection{Elemental Abundances}
\begin{figure*}
    \includegraphics[width=\textwidth]{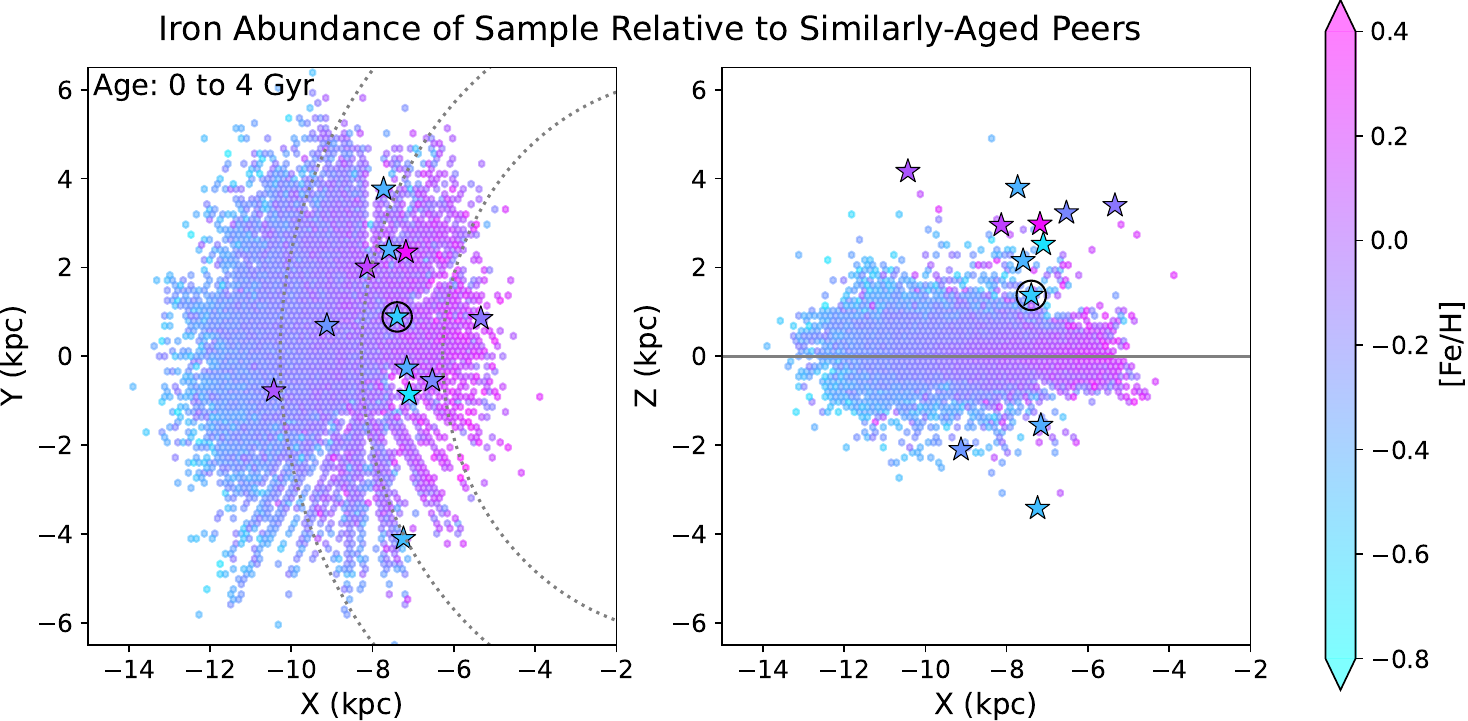}
    \caption{The x-y (left) and x-z (right) distribution of all stars with thin disk abundances for age estimates up to four billion years, with the slow and retrograde stars in this age range over-plotted as a star-shaped marker. Each hexagon is color-coded by its mean iron abundance, while each slow and retrograde star is colored by its iron abundance. Stars within 1.5 kpc of the mid-plane  (i.e. five thin disk scale heights, e.g. \citealt{bland_hawthorn_2016}) are circled. The dotted gray lines are circles with radii equal to the Solar Galactocentric radius ($R_\odot$) and $R_\odot \pm 2$~kpc, each centered on the Galactic center.}
    \label{fig:spatial_0_to_4}
\end{figure*}

\begin{figure*}
    \includegraphics[width=\textwidth]{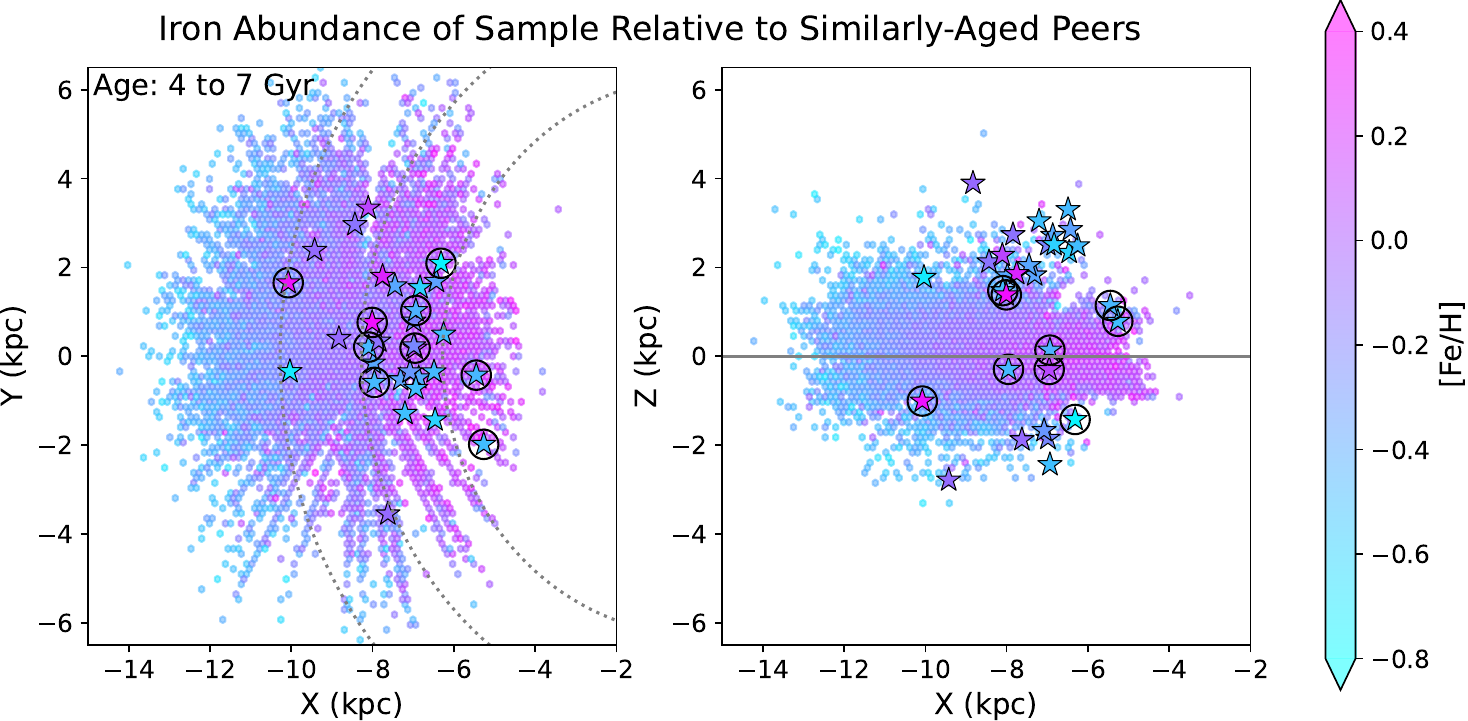}
    \caption{The same as Figure \ref{fig:spatial_0_to_4}, but for ages between four and seven billion years.}
    \label{fig:spatial_4_to_7}
\end{figure*}

\begin{figure*}
    \includegraphics[width=\textwidth]{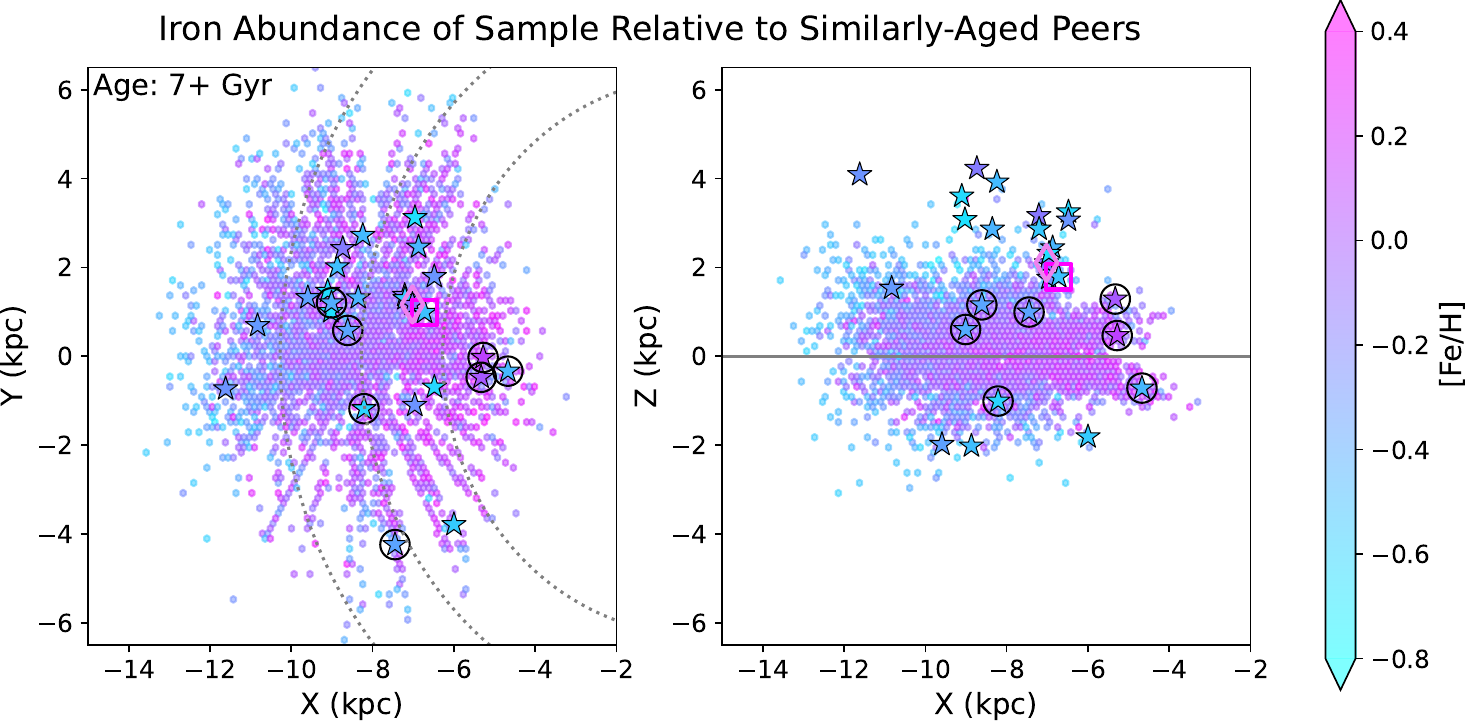}
    \caption{The same as Figure \ref{fig:spatial_0_to_4}, but for ages above seven billion years.}
    \label{fig:spatial_7_plus}
\end{figure*}

The Milky Way is known to have vertical and radial abundance gradients, and the slope(s) of the gradient(s) depend on stellar age (see e.g. \citealt{anders_2017}). Stars that have had their orbital radii altered are thus likely to have abundances (namely iron and alpha-abundances) that differ from the average abundances of similarly-aged stars at the same location in the disk. We split the data into wide age bins that roughly trace \lq younger' ($<$ 4 Gyr), \lq intermediate age' (4 - 7 Gyr), and \lq old' (7+ Gyr) stars to compare the stars in the sample to their similarly aged thin disk peers. The spatial distributions of all thin disk stars in these age bins are shown in Figures \ref{fig:spatial_0_to_4} to \ref{fig:spatial_7_plus}, color-coded by the mean iron abundance in a given hexagonal pixel. The slow and retrograde stars in each age bin are over-plotted and color-coded by their individual iron abundances. Some of these stars are relatively more metal-poor than their peers (i.e. bluer) at a given location, indicating that they may have come from larger radii or larger distances from the plane. Other stars in the sample are more metal-rich (i.e. pinker), and they may have come from smaller radii and/or smaller heights above the plane. These differences relative to similarly aged peers are an expected result of many of the mechanisms that we discuss in Section \ref{sec:mechanisms}.

\begin{figure*}
    \includegraphics[width=\textwidth]{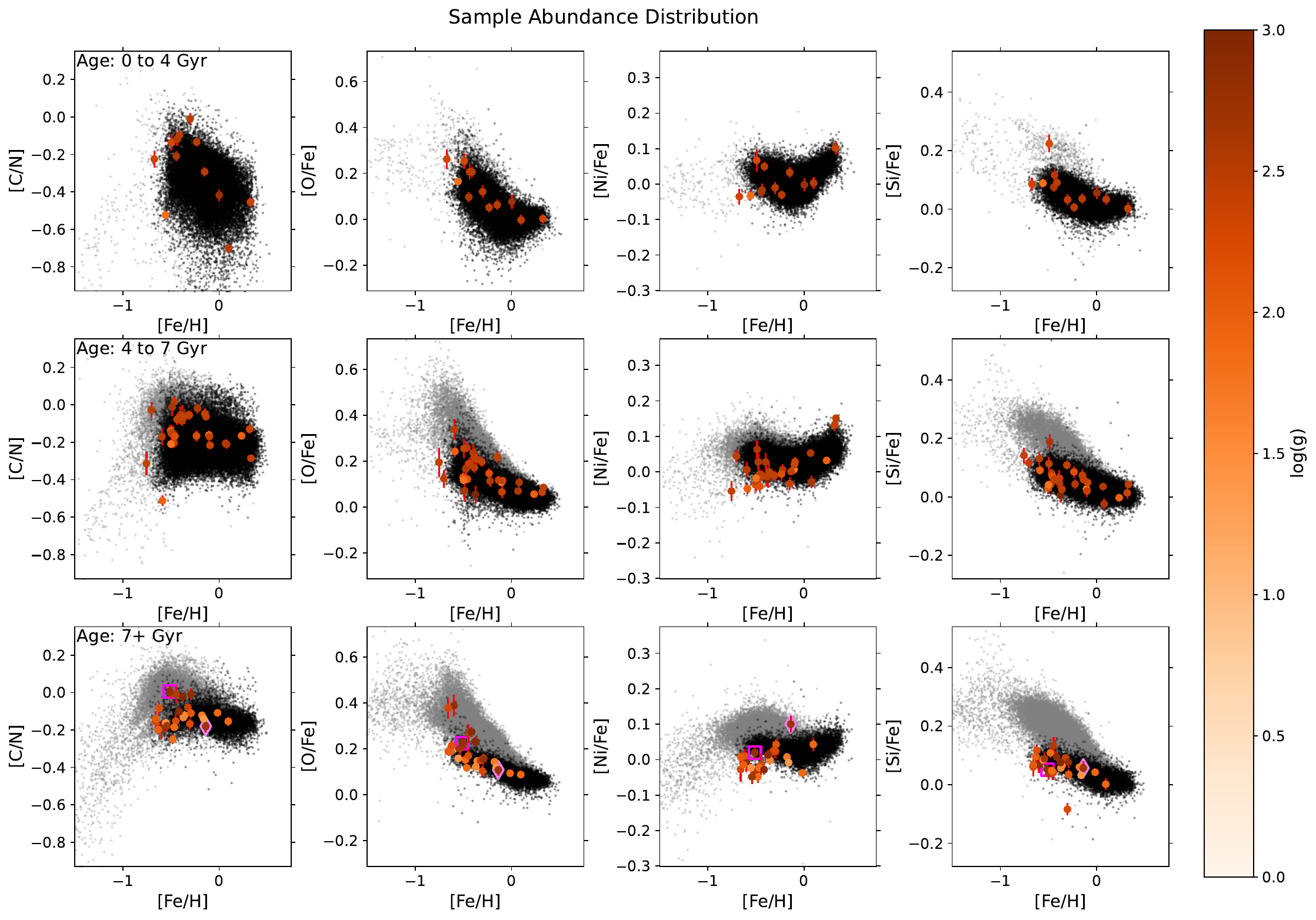}
    \caption{The abundances of each of the \lq most reliable' elements in APOGEE for all stars (grey), the thin disk (black), and the slow and retrograde sample (larger points color-coded by log(g)), each split into age bins Error bars are the errors provided in APOGEE, and for [C/N] the errors for [C/Fe] and [N/Fe] are added in quadrature. The slow and retrograde sample has been split into three rough age bins for clarity.}
    \label{fig:abundances_reliable}
\end{figure*}

We next investigate how individual elemental abundance estimates of the sample compare to the broader Milky Way. Figure \ref{fig:abundances_reliable} shows the slow and retrograde sample along with all thin disk stars (black) over-plotted on all quality APOGEE stars (grey) for a selected set of abundances. The sample is split into three separate rows --- again by age bins --- for clarity. The slow and retrograde sample is over-plotted as larger symbols and color-coded by log(g). The abundances presented in Figure \ref{fig:abundances_reliable} include only those abundances selected to be most reliable in the APOGEE DR17 ASPCAP documentation, excluding [Mg/Fe], [Mn/Fe], and [Al/Fe]. The remaining elements include alpha elements (oxygen, silicon), an iron peak element (nickel), and light alpha elements (carbon and nitrogen). As the surface abundances of carbon and nitrogen in red giant branch stars evolve during the first dredge-up in a way that is mass (and therefore age) dependent \citep[see e.g.][]{iben1965,iben_1967}, we opt to plot [Fe/H] vs [C/N]. The general age trend for [Fe/H] versus [C/N] is such that older stars tend to have higher [C/N], while younger stars will have lower [C/N] (see e.g. \citealt{martig_2016} and references therein). There are a number of stars outside of the bulk of the [C/N] distribution, some with high [C/N] and others with very low [C/N] at a given [Fe/H]. The carbon to nitrogen ratios in the atmospheres of these stars may have been altered by other processes beyond the dredge-up (e.g. binary interaction, \citealt{izzard_2018}), and their abundances should likely not be taken as robust age indicators.

Figure \ref{fig:abundances_reliable} reveals that there are some stars in the sample that are outliers in other abundance spaces. Some slow and retrograde stars are oxygen-enhanced relative to the bulk of the thin disk population, with little to no log(g) differences (as indicated with the color-coding). Given that the sample is selected to have thin-disk-like magnesium abundances, one may expect that the other alpha abundances would also be within the bulk of the thin disk distribution. However, the stars with the highest [O/Fe] abundances also have relatively large errors, reducing the possible significance of their enhancement. The same is true for the high [Si/Fe] stars. Further, while oxygen and magnesium are both alpha elements, these elements do not necessarily exactly trace one another (as shown in e.g. \citealt{franchini_2021} for dwarf stars).

Finally, we note that relative to the total thin disk sample, the over-all distribution of slow and retrograde stars is somewhat skewed towards the lower [Mg/H] and [Fe/H] end of Figure 1 and 5, respectively. There are a number of potential sources for this skew (e.g. selection effects, physical origins such as the different levels of contributions from the various mechanisms discussed here, et cetera), making it difficult to interpret. We return to elemental abundances when linking individual stars to mechanisms.

\subsection{Comparison to Known Chemo-Dynamical Substructures}\label{section:substructure_comp}

\begin{figure*}
    \includegraphics[width=\textwidth]{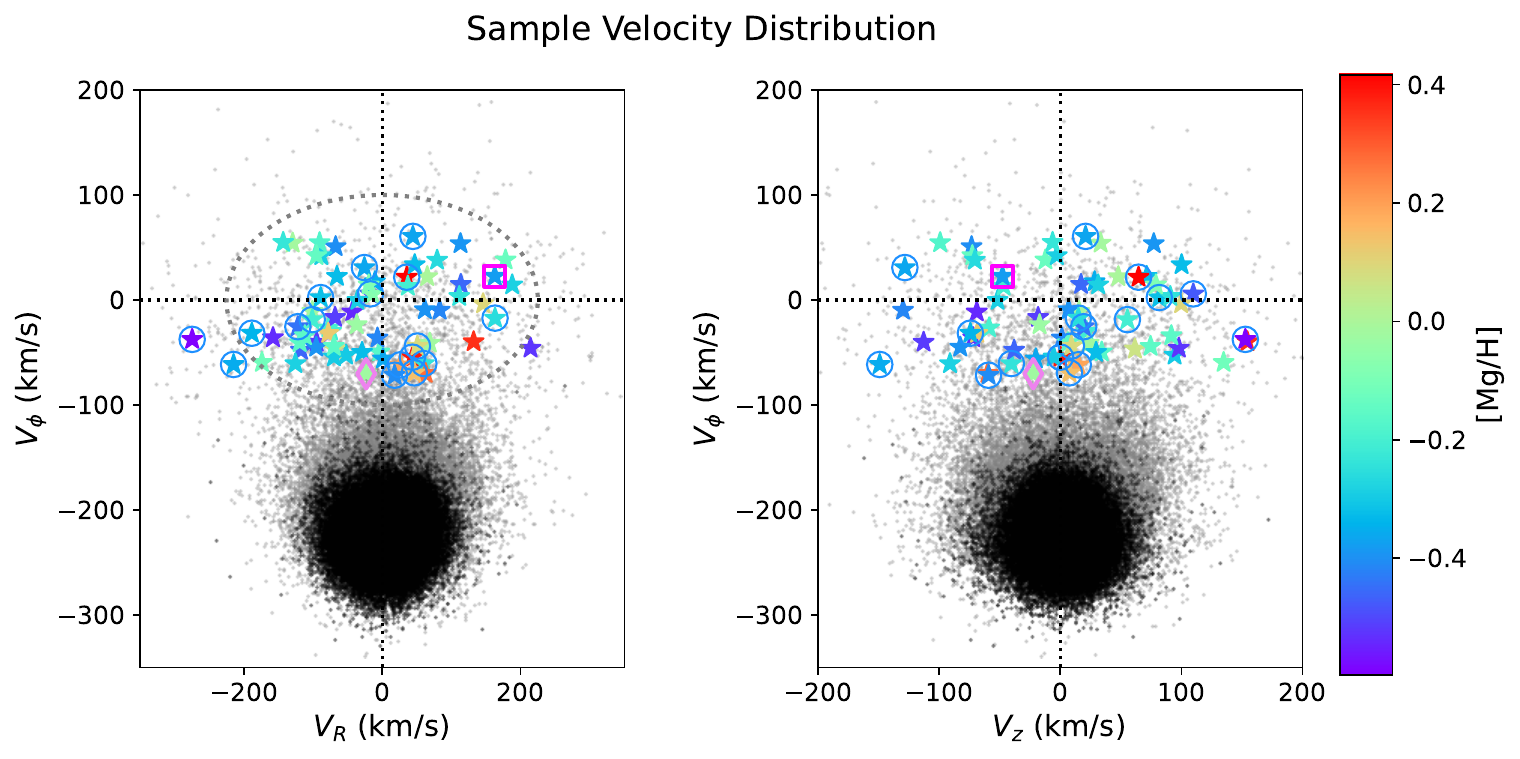}
    \caption{The velocity distribution of slow and retrograde stars with thin disk abundances (color-coded by [Mg/H]), along with the full APOGEE sample (grey) and stars with thin-disk abundances (black). Stars in blue circles are those within 1.5 kpc of the Galactic mid-plane. Left: the distribution in the $V_R$ -- $V_\phi$ plane, alongside an oval roughly indicating the typical location of GSE stars. Right: the distribution in the $V_z$ -- $V_\phi$ plane.}
   \label{fig:velocity_dist}
\end{figure*}

The velocity distribution of the sample compared to all quality stars (grey) and thin disk stars (black) is shown in Figure \ref{fig:velocity_dist}. As anticipated, the slow and retrograde stars are offset from the locus of the thin disk stars. The region that the slow and retrograde stars occupy is also home to chemo-dynamical (sub)structures. In particular, the debris from the \lq Gaia Sausage Enceladus' (GSE, \citealt{belokurov_2018}, \citealt{helmi_2018}) progenitor, proposed to be the Milky Way's last major merger, has been shown to occupy a similar swath of $V_{R}$ -- $V_{\phi}$ space, with stars having low or retrograde angular momentum. This overlap should not be surprising, however, given that we select stars with slow or retrograde $V_\phi$ velocities. Despite the kinematic overlap between the slow and retrograde sample and the accreted GSE population, these populations should have different abundance distributions, allowing us to separate the two. The way that GSE candidate stars are selected will influence the resulting abundance distribution (see Section \ref{sec:discussion}), but as seen in e.g. \cite{carrillo_2024}, there are few to no stars with thin disk abundances for a number of choices of GSE selection cuts. Indeed, we find that there is minimal or no simultaneous overlap in abundances and kinematics with the numerous substructures presented in \cite{horta_2023}, regardless of accreted or in-situ origin. There is also no kinematical overlap with the Anticenter Stream or the Monoceros ring, both of which are structures with likely disk origins that have $V_\phi$ similar to the rotation curve (see e.g. velocity information in \citealt{deboer_2018}).

There is plausibly overlap with the likely in-situ \textit{Eos} substructure identified in \cite{myeong_2022}, which is comprised of stars on eccentric orbits with an abundance distribution that overlaps with the thick disk and metal-poor end of the thin disk. However, our slow and retrograde sample extends more iron- and [Mg/Mn]-rich than the distribution of candidate \textit{Eos} stars, limiting the overlap between the samples. Similarly, it is possible that our sample contains stars that have been associated with the low angular momentum and/or retrograde \textit{Splash} component of the Milky Way (see \citealt{belokurov_2020}). Given the rather non-restrictive cuts in [Fe/H] and $V_\phi$ that are used to identify \textit{Splash} stars, this overlap is anticipated. Stars belonging to the \textit{Splash} generally have abundances that are more consistent with the high-alpha disk of the Milky Way than the thin disk, though a low-alpha component of the \textit{Splash} has also been explored \citep[e.g.][]{zhao_2021,lee_2023, nepal_2024}. It remains unclear if this low-alpha component formed entirely in-situ (as suggested in \citealt{zhao_2021}, who invoke the clumpy disk scenario presented in \citealt{amarante_2020} in which a major merger is not required) or if it is a combination of accreted stars and stars that formed in a burst following the last major merger event (as suggested in \citealt{lee_2023}). \textit{Eos} may itself be a part of the low-alpha component of the \textit{Splash} \citep{lee_2023}

The possible overlap with \textit{Eos} and the \textit{Splash} is interesting in its own right, as the origin of these substructures remains debated. Further, while some or most stars with abundances and kinematics consistent with \textit{Eos} or the low-alpha component of the \textit{Splash} may have originated from one of the scenarios outlined above, some stars --- especially the more metal-rich stars in our sample --- may not have. Indeed, \cite{bonaca_2017} discuss a variety of possible origins for metal-rich halo stars (predating the \citealt{belokurov_2020} \textit{Splash} analysis), including the possibility that some small fraction are runaway stars from the disk. Our analysis is complementary to this earlier work and explores the idea of thin disk stars on halo-like orbits. We assume a thin disk origin for slow and retrograde stars and investigate possible mechanisms for altering stellar orbits in Section \ref{sec:mechanisms}, then in Section \ref{sec:discussion}, discuss alternate halo origins.

\section{Mechanisms}\label{sec:mechanisms}
Without a clear association with known chemo-dynamical substructures, we investigate a variety of plausible mechanisms that can alter thin-disk-like orbits to produce slow or retrograde velocities. This section operates with the standard assumption that stars born with thin-disk-like abundances are also born on thin-disk-like orbits. Three of these mechanisms --- dynamical ejection from a stellar cluster, supernova ejection from a binary, and ejection from a stellar binary interaction with the central supermassive black hole (often called the Hills mechanism, \citealt{hills_1988}) --- are proposed mechanisms for producing runaway or high velocity stars. Indeed, a mechanism that can create a runaway star should also be able to produce a low-$V_\phi$ or retrograde star, provided there is no preferential ejection angle. The majority of stars ejected from both the cluster ejection and binary supernova mechanisms have velocities much lower than $\sim 200$ km~s$^{-1}$ (see e.g. \citealt{perets_subr_2012}, \citealt{renzo_2019}), and ejection velocities fast enough to produce slow or retrograde stars are rare. Nevertheless, it is plausible that such mechanisms could produce low-$V_\phi$ or retrograde stars, which are themselves rare.

Analyses of runaway or high-velocity star ejection mechanisms tend to focus on ejected main sequence stars of relatively high mass \citep[i.e. blue stars,][]{hvs_review}. This focus has emerged for a variety of reasons. For example, searches for high velocity stars can be optimized by searching for blue stars in the stellar halo (see e.g. discussion in \citealt{kollmeier_2007}). The stars in our analysis are red giant stars. The time that a star spends in the main sequence stage is typically longer than the time that it spends as a red giant, and we assume that the stars in the sample were most likely ejected as main sequence stars and have since evolved to be giants. This assumption also avoids possible issues concerning the ejection of red giants, as some processes may impact giants differently than main sequence stars (e.g. close binary evolution). The relative shortness of the red giant phase also means that ejected stars in the giant phase are more rare than those in the (lower-mass) main sequence phase, however, this reduction in probability should apply to all mechanisms and thus is unlikely to change our conclusions. In the following subsections, we estimate the rate of production of slow and retrograde stars via these three mechanisms, and restrict these estimates to stellar mass ranges reasonable for red giants that are at least a few gigayears old.

The remaining plausible mechanisms for producing slow and retrograde stars are ejection or stripping from a satellite galaxy, scattering by dense star-forming clumps, interactions with the Galactic bar, and torque/scattering from a passing satellite. Unlike the previous mechanisms, there is no preference for the stellar type or mass typically assumed in these scenarios, though the majority of analyses of these mechanisms use simulations with massive star `particles' due to computational limitations. Given that we select only stars with evolved, thin-disk-like abundances, we can already reject ejection or stripping from a satellite galaxy (though see Section \ref{sec:discussion} for further discussion). We investigate these remaining possibilities below.

\subsection{Dynamical Ejection from a Cluster}

In the dynamical ejection mechanism scenario, a star is ejected from a cluster due to multi-body interactions (see e.g. \citealt{poveda_1967}). These stars are ejected from their clusters after experiencing a chance encounter, typically as part of or with a multiple system, that imparts a sufficiently high velocity kick. There are a number of different kinds of interactions that can cause dynamical ejection (see e.g. the recent summary in the context of globular clusters presented in \citealt{newlin_2023} and references therein), but here we are only interested in the net effect of the interactions. Namely, we are interested in the distribution of ejection velocities from stellar clusters, as well as the typical mass distribution of the ejected stars. In the multi-body interaction case, the kick velocities depend on the properties of the binary distribution, as the magnitude of the kick is controlled by the binding energy of the binary, while the rate of ejections is sensitive to other initial condition parameters \citep[see e.g. the ranges explored in][]{perets_subr_2012, oh_2016}. Higher velocity ejections tend to come from denser, more massive clusters (e.g. \citealt{perets_subr_2012}). For simplicity we assume that the clusters are co-orbiting with the disk, with $|V_\phi| \sim V_c$.

We compute a rough estimate for the rate at which slow and retrograde stars are produced via dynamical ejection from dense/massive clusters, such as globular clusters or similar sites of star formation in the Milky Way. Our computation closely follows that given in \cite{hvs_review} for hyper-runaway stars and relies on the results of the simulations presented in \cite{perets_subr_2012} and \cite{newlin_2023}. We restrict our estimates to the well-studied regime of young clusters (as in \citealt{perets_subr_2012}) and consider only the earliest times of the \cite{newlin_2023} simulations. We briefly justify this choice here. The globular clusters observed in the Milky Way today are typically old and their ejection history can be complicated. For example, as shown in (e.g.) \cite{newlin_2023}, the dominant ejection mechanism within globular clusters (e.g. binary-single interaction, three-body binary formation, et cetera) can change as a function of time, and the ejection rate declines as the globular cluster ages. In addition, the \cite{perets_subr_2012} models do not include stellar evolution and the simulations stop before the first supernovae would occur, providing estimates only for ejections by dynamical encounters between stars/stellar systems early in the cluster history. We assume that the simulation-based estimates are representative of the early epochs of the known globular clusters, as well as any now-dissolved birth clusters, such as dense open clusters \citep[see e.g.,][who find that the young Orion nebula cluster could produce sufficiently fast ejections]{schoettler_2020}.

The smallest required velocity for a star on a disk-like orbit to become slow or retrograde occurs when the star is ejected at $\sim 150$ km~s$^{-1}$ exactly, or nearly exactly, in the direction opposite the sense of disk rotation. For ejection directions that are randomly oriented, we adopt the assumption given in \cite{hvs_review} that $\sim 10\%$ of ejections will be in a given direction (in their case, aligned with disk rotation to boost velocities). Higher total velocities are required if the star is ejected at an angle relative to the disk rotation. The fraction of dynamically-ejected runaways at a given velocity decreases as a function of velocity \citep[see e.g.][]{perets_subr_2012,hvs_review}, which can negate gains in the total rate made through the increased range of acceptable ejection angles, so we here only estimate the rate for the minimum required velocity. Finally, we note that stars ejected from clusters could reach large vertical orbital extents, given an adequate combination of velocity and ejection angle. 

Inspection of Figure 1 of \cite{perets_subr_2012} indicates that the fraction of runaways ejected at $V \sim 150$ km~s$^{-1}$ is $\sim 2 \times 10^{-2}$ for their model 009. Combining this number with an estimate for the rate of stars ejected from very young globular clusters via stellar multi-body interactions --- $2.4 \times10^{-3}$ stars per year \citep[][Weatherford, priv. comm.]{newlin_2023} --- and that $10\%$ of stars are ejected in the direction opposite that of disk rotation, we find that $(0.1) \times (2 \times 10^{-2}) \times (2.4 \times 10^{-3} \rm{ stars}$ $\rm{ yr}^{-1}) \sim 4.8 \times 10^{-6}$ stars are ejected per year with this combination of velocity and direction. Turning to Figure 4 of \cite{perets_subr_2012}, we see that $\sim 1\%$ of stars ejected with $\rm{V} > 200$ km~s$^{-1}$ and $\sim 5\%$ of stars ejected with $\rm{V} > 100$ km~s$^{-1}$ have mass between $\sim 1 - 1.5 \rm{M}_\odot$, averaging these values indicates that there are $\sim 1.44 \times 10^{-7}$ lower mass, minimum velocity ejections per year. If this rate has remained constant for $\sim 9$ Gyr (i.e. since at least the approximate birth time of the second oldest star in the sample), there should be $\sim 1,300$ slow stars in the Milky Way disk produced via this mechanism, some subset of which could be in the APOGEE sample. Note that a constant rate of ejection would assume that new clusters are forming and dissolving over time, or that the ejection rate from globular clusters remains approximately constant as they age. As mentioned above, the rate of ejections from globular clusters drops significantly over the first gigayear of the clsuters' life (see Figure 3 of \citealt{newlin_2023}), but favorable Galactic orbits would lower the minimum velocity threshold and may help to somewhat compensate for this lower rate.

\subsubsection{Possible Observational Signatures}

If stars born in the same cluster are homogeneous in their elemental abundances and cluster abundances are distinct from one another, with sufficient data it should be possible to identify the birth cluster of a given star (this is the goal of strong \lq chemical tagging', see e.g. \citealt{freeman_2002}). While it is unclear the extent to which strong chemical tagging is possible, it is true that a star ejected from a cluster should have chemistry consistent with the rest of the cluster, barring changes in surface abundances due to dredge-up(s), binary interactions, et cetera. We consider abundances consistent with a given globular cluster to be a sign that a given star was possibly ejected from that cluster, and we do not attempt to trace stars back to natal clusters that have since dissolved. As globular clusters are typically old (10+ Gyr), it is likely that the Galactic potential has changed significantly since the clusters formed. We therefore do not attempt to dynamically link stars to clusters via backward integration of orbits. Finally, we note that the unusual abundances of globular clusters (see below) may negatively influence the accuracy of spectrum-based age estimates for cluster members, and we do not consider age information in this comparison.

\subsubsection{Identification of Candidate Stars}

 We investigate the properties of clusters within the APOGEE globular cluster value-added catalog \citep{schiavon_2024}, considering only stars with a line-of-sight velocity membership probability above 50$\%$ (\texttt{RV\_Prob}$> 0.5$). We enforce the same thin disk abundance selection and spectroscopy quality cuts listed Section \ref{sec:selection}, this time ignoring astrometric quality cuts and allowing stars have any value of \texttt{MEMBERFLAG}. We also allow bits other than 2 and 4 of \texttt{EXTRATARG} to be set. Most globular clusters are metal-poor, and are thus excluded from our analysis. Globular clusters often have element anti-correlations (see e.g. the APOGEE-based analysis in \citealt{meszaros_2020}) in aluminum-magnesium, nitrogen-carbon, and/or sodium-oxygen, leading to some stars being relatively enhanced (depleted) in aluminum (sodium), nitrogen (carbon) and/or sodium (oxygen). Our sample selection in the [Al/Fe] vs [Mg/Mn] plane will likely exclude second generation of aluminum-rich cluster stars ([Al/Fe] $> 0.3$ in \citealt{meszaros_2020}). Some clusters may therefore be excluded if the APOGEE observations contain few or no first-generation stars. \cite{meszaros_2020} also show that clusters with [Fe/H] $\gtrsim -1$ -- such as NGC 6388 -- tend to have only a first generation population in aluminum/magnesium, but can still have first and second generation populations in carbon/nitrogen. As such, our selection in the [Al/Fe] vs [Mg/Mn] plane does not necessarily preclude a spread in [C/N] from cluster members.

\begin{figure*}
    \includegraphics[width=\textwidth]{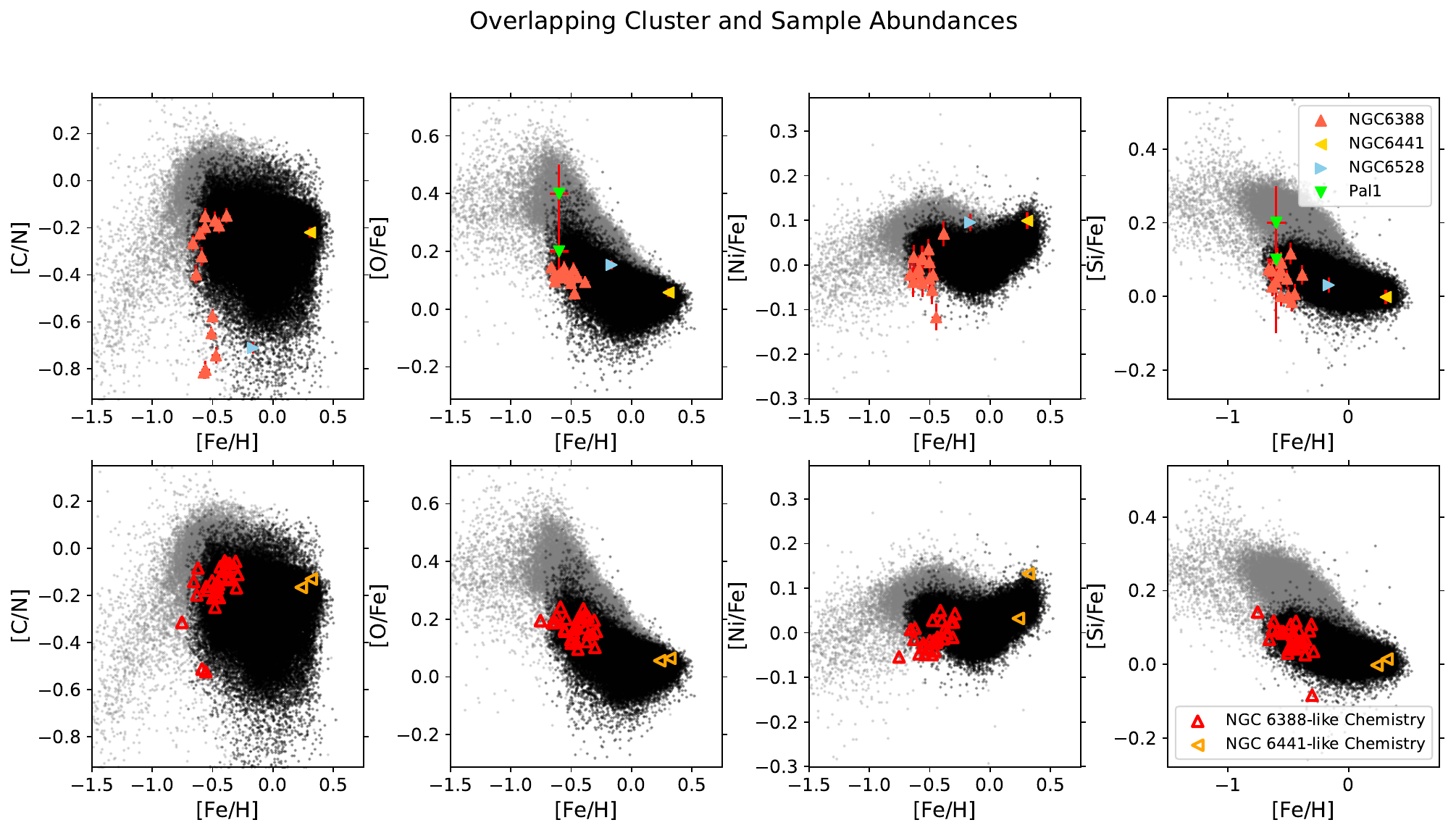}
    \caption{The elemental abundances of the slow and retrograde stars that are plausibly linked to globular clusters (bottom row, colored open triangles, excluding Pal1) alongside those clusters (top row, filled colored triangles), the thin disk (black dots) and all quality APOGEE stars (grey dots). Stars with thin-disk-like abundances that are identified cluster members are shown as red, orange, blue, and green triangles with different orientations in the top panels, and slow and retrograde stars with abundances similar to a given cluster (i.e. within $\pm 0.1$ the maximum and minimum values, respectively, of the abundances of the member stars with thin-disk-like abundances) are shown with open triangles of the respective color and orientation in the bottom panels. Errors for Pal1 are those from \cite{jahandar_2017}, while the error bars for all other cluster stars are those quoted in APOGEE. Carbon-to-nitrogen error bars are the quoted errors added in quadrature. The lowest [C/N] ratios of NGC6388 are below the y-axis limits.}
    \label{fig:simchem}
\end{figure*}

There are four clusters that have member stars that pass the cuts given above: NGC 6388, NGC 6441, NGC 6528, and Palomar 1 (Pal1) \footnote{\cite{jahandar_2017} note that the two stars in APOGEE DR12 flagged as likely members of Pal1 have persistence issues in their spectra that impact the derived abundances. The APOGEE DR17 parameters for these stars are quite similar to the DR12 parameters, and in this figure we adopt the values from the re-analysis for a subset of elements of these stars presented in \cite{jahandar_2017}.}, and we present the elemental abundances of the member stars that pass these cuts in Figure \ref{fig:simchem}. Over the years, numerous studies have classified globular clusters as belonging to different components of the Milky Way, such as the bulge or the disk \citep[see e.g. the recent analyses presented in in][]{massari_2019,horta_2020,callingham_2022}. For example, the chemo-dynamical classifications presented in \cite{callingham_2022} assign NGC 6388 and NGC 6528 as belonging to the bulge, NGC 6441 as belonging to the disk, and Pal 1 as \lq ungrouped’. While it is not the same as a disk origin, a bulge provenance for some cluster-ejected stars is still consistent with the assumption that the slow and retrograde stars did not originate in the halo. Finally, given that orbits can evolve, the present-day dynamical classification of an in-situ cluster does not necessarily indicate the dynamical classification at birth or other earlier epochs.


As seen in \cite{bajkova_2022}, the current orbits of these clusters reach a range of vertical extents, and not all of the clusters have $V_\phi$ velocities close to the $-230$ km~s$^{-1}$ assumed throughout this analysis (in fact, NGC 6388 has a retrograde velocity). Slower $V_\phi$ velocities make it easier to produce slow and retrograde stars by lowering the required ejection velocity, and larger heights above the plane would make it easier to produce stars at large heights from the plane (see e.g. Figures \ref{fig:spatial_0_to_4} to \ref{fig:spatial_7_plus}). On the present orbits with the ejection rate given above, these clusters could have produced more slow and retrograde stars than anticipated from our order-of-magnitude estimates.

 We identify stars that are plausibly linked to a given cluster as those with abundances within $\pm 0.1$ dex of the maximum and minimum values, respectively, of the abundances of the cluster members with thin-disk-like abundances. We make this selection using the abundances shown in Figure \ref{fig:simchem}, as well as [Mg/Fe], [Mn/Fe], [Al/Fe], and [Fe/H]. We omit Pal1 due to the large error bars and the limited elements obtained in the reanalysis presented in \citealt{jahandar_2017}. We present the abundances of the stars plausibly linked to these clusters in Figure \ref{fig:simchem}, alongside those of their possible parent clusters as well as the thin disk and quality star background. There are 30 stars similar to NGC 6388 and 2 stars similar to NGC 6441, while the remaining stars do not appear to be similar in abundances to the remaining globular clusters. We consider these 32 stars to be consistent with this mechanism. However, we cannot rule out the possibility that any given slow or retrograde star was ejected from a birth cluster that has since dissolved. It may also be possible that some stars in the sample are consistent with an unobserved or underrepresented first generation population in the APOGEE clusters. As such, we consider dynamical ejection to be a possible mechanism for all stars in the sample.

\subsection{Binary Supernova Companion}
In the binary supernova ejection mechanism, the massive primary star of a binary system undergoes core collapse and causes the secondary star in the system to be ejected (e.g. \citealt{blaauw_1961}). Ejection is not necessarily guaranteed, and requires that the mass loss and/or supernova kick are sufficiently high to unbind the binary (see e.g. \citealt{renzo_2019}). Single-degenerate scenario Type Ia supernovae may also eject (stripped) main sequence stars. However, it is uncertain how much the single-degenerate scenario channel contributes to the total type Ia supernova rate (see e.g. the review presented in \citealt{liu_review}), and to what extent main sequence stars in the mass range of interest contribute to the distribution of non-degenerate companions. We thus focus on core-collapse supernovae.

We assume the same initial velocities as in the dynamical ejection scenario, such that for the ejected star to be retrograde, it must be ejected with velocity $\gtrsim 150$ km~s$^{-1}$ in the direction opposite that of disk rotation. The distribution of ejection velocities is highly dependent on the (uncertain) distributions of both the supernovae kick velocities and binary properties before to the supernova explosion (see e.g. \citealt{renzo_2019}, \citealt{evans_2020} and references therein). We again turn to simulations to estimate the rate of retrograde star production.

\cite{evans_2020} run suites of binary star synthesis simulations to estimate the per-system probability of high velocity ejections. These suites test a range of input parameter assumptions, and we employ the simulation with the fiducial estimates for the input parameters (matching those assumed in \citealt{renzo_2019}). We use their publicly-available simulation outputs and follow their assumptions in the latter portion of their Section 2.4. Namely,  we assume a minimum zero-age main sequence mass for core collapse progenitors of $8 \rm{M}_\odot$, and that all stars above this limit will undergo core collapse. We then compute the fraction of the systems with primaries above $8 \rm{M}_\odot$ that eject secondaries with mass between $1$ and $1.5 \rm{M}_\odot$ post core-collapse and ejection velocities $V > 150$ km~s$^{-1}$, finding $4.4 \times 10^{-4}$. Multiplying this estimate with the rate of core collapse supernovae per year \citep[$\sim 5 \times 10^{-3}$ supernovae yr$^{-1}$][]{quintana_2025} and our estimate that $\sim 10\%$ of stars are ejected opposite the direction of disk rotation, we find that there are $2.95 \times 10^{-7}$ ejections per year that could produce slow and retrograde stars. If we again assume that this rate has remained constant for 9 Gyr, there would be $\sim 2,000$ slow stars produced via this mechanism. Note that this number drops off quickly with increasing velocity: at $V > 200$ km~s$^{-1}$, for example, there would be $\lesssim 200$ slow or retrograde stars produced. As with dynamical ejection above, stars ejected from binary supernovae could reach large vertical orbital extents, given an adequate combination of velocity and ejection angle.

\subsubsection{Possible Observational Signatures}
Depending on the evolution of the binary pair, the ejected secondary may display signs typical of binary interaction (as discussed in e.g. \citealt{renzo_2019}). If the ejected secondary star gained mass, for example, it could rotate relatively rapidly for its age and have a magnetic field (see e.g. \citealt{yu_2024} and references therein). Binary interactions can also push the surface carbon-to-nitrogen ratios to lower values \citep[e.g.][]{izzard_2018}. Further, if the secondary star accreted elements fused in the primary or in the supernova, it could be enhanced in certain elements, such as the alpha elements magnesium and silicon (as suggested in \citealt{israelian_1999}, see also the reanalysis in \citealt{gonzalez_2008} and references therein). If enrichment in alpha elements is too high, however, it could remove the secondary from the sample considered here. It is also possible that there are no clear signs in the elemental abundances for some supernova-ejected stars, depending on the mixing timescale (see e.g. the discussion in \citealt{liu_2015}). Finally, given that the secondary is ejected from its primary, the secondary likely would show no kinematic indicators of binarity (e.g. line-of-sight velocity variability), as this would require an unlikely tight binary within a trinary where the third massive companion goes supernova.

\subsubsection{Identification of Candidate Stars}
As seen in Figure \ref{fig:abundances_reliable}, there are a number of stars with [C/N] ratios below the bulk of the population. These low [C/N] abundances could be indicators of past binary interaction(s) with a supernova companion, and we consider stars with low ratios ([C/N] $<$ -0.4 dex) to be consistent with this mechanism. Only one star in the slow and retrograde sample has an estimated rotation velocity. The reported velocity establishes this star as a rapid rotator ($v\sin(i) \sim 20$ km~s$^{-1}$, where the typical rapid rotation threshold is $\sim 10$ km~s$^{-1}$, see e.g. discussion in \citealt{tayar_2015}), and we consider this star to be consistent with this mechanism. This star (indicated with an open violet diamond in the figures) does not appear to be an abundance outlier relative to the thin disk population. Finally, the star with high velocity variability is likely a binary and thus unlikely to be related to this mechanism. For all other stars, given that there may not be a abundance signature from supernova ejection, we consider binary supernova ejection to be plausible.

\eject
\subsection{Ejection from the Central Supermassive Black Hole}

In the ejection from the central black hole scenario (i.e. Hills mechanism), a star is ejected from a stellar binary when the binary is disrupted by the supermassive black hole at the center of the Galaxy (see e.g. \citealt{hills_1988}). There are a few variations of this mechanism, for example the case of a single star interacting with binary black holes (e.g. \citealt{yu_2003}), but here we focus on the \lq classic' binary star and single, supermassive black hole scenario. While the Hills mechanism is often discussed in the context of very high velocity stars, it does produce a velocity distribution that includes values below the local escape velocity (see e.g. \citealt{bromley_2006}). Given the velocities of the stars in the sample, we focus on these lower velocity ejections. The ejection velocities depend on the binary properties and interaction geometry, and the final observed velocity at a given location depends on both this ejection velocity and the Galactic potential. Stars ejected via the Hills mechanism are on radial orbits and will thus generally have low amplitude $\rm{V}_\phi$ velocities. We seek to estimate only the rate of production of bound stars over a reasonable stellar mass range that can travel to the distance range spanned by the sample.

We turn to the analysis presented in \cite{kenyon_2014} to estimate the rate of production of bound stars. Those authors simulate ejection via the Hills mechanism and integrate the resulting orbits in a Milky Way-like gravitational potential. They find that $5\%$ of ejected stars have sufficient velocities to reach distances of $5$ to $20$ kpc from the Galactic center while remaining bound to the Galaxy, and further estimate that $\sim 10\%$ of ejections have mass reasonable for a G-type star ($\sim 0.8$ - $1.2 \rm{M}_\odot$ --- note that this mass range is not identical to those considered above). If we adopt an ejection rate of $\sim 1 \times 10^{-4}$ stars ejected from the central black hole per year (see e.g. \citealt{hvs_review}), we get a final rate of $(0.05) \times (0.10) \times (1 \times 10^{-4} $ ejections yr$^{-1}) = 5.0 \times 10^{-7}$ ejections per year. If this rate has remained constant for 9 Gyr, there should be $\sim 4,500$ slow and retrograde stars of reasonable mass. It is worth noting here that while a rate of $\sim 10^{-4}$ per year is commonly assumed, some more recent analyses have found ejection rates closer to $\sim 10^{-5}$ per year \citep[e.g.][]{marchetti_2022,sill_2025}, which would lower this final estimate by up to an order of magnitude.

\subsubsection{Possible Observational Signatures}

Stars ejected from the central supermassive black hole most likely formed within the central few kiloparsecs of the Galaxy. While it is possible for stars that formed at larger radii to be moved towards the black hole via dynamical processes, given the surface density profile of the disk these stars should be more rare than the stars born locally. The metallicity gradient in the disk is such that stars formed at smaller radii are more metal-rich. Solar or super-Solar metallicity is thus a possible sign that a star could have been formed near and ejected from the Galactic center. Indeed, the most likely candidate to date of a Hills mechanism-ejected star has super-Solar iron abundance, see \cite{koposov_2020}. At most radii, such ejected stars would have higher metallicities than their peers, as they will have moved from their (smaller) birth radii outwards to larger radii. In the absence of any orbital evolution, Hills mechanism ejected stars should be on radial orbits. However, given that orbits can be altered by (e.g.) the bar, we loosen this constraint, and consider high eccentricity to be a potential signature of this mechanism. These orbits can also reach large vertical extents, depending on ejection angle. Finally, unlike most hyper-velocity star searches, the stars could be outbound from the Galactic center or inbound and returning from a previous orbit.

\subsubsection{Identification of Candidate Stars}
There are nine stars with [Fe/H] $> 0$ dex. The ratio of radial to azimuthal velocities for each of these stars indicate that they are on eccentric, rather radial orbits, with the minimum $|\frac{\rm{V}_R}{\rm{V}_\phi}|$ of this sample being greater than $\frac{1}{3}$. These nine stars are the most consistent candidates in the sample for the Hills ejection mechanism. As the orbit of the star after ejection depends on the Galactic potential and could be influenced by the Galactic bar, we do not attempt to integrate orbits to identify those that may have come from the Galactic center.

\subsection{Scattering in the Early Clumpy Disk}
Isolated galaxy simulations featuring early, dense star forming clumps have had success in producing thin disk, thick disk, and \textit{Splash}-like populations (see e.g. \citealt{clarke_2019}, \citealt{amarante_2020}, \citealt{leandro}), and in-situ retrograde populations have been investigated in some of these simulations. In these simulations, retrograde stars are produced by scattering off of the early clumps, and both high- and low-alpha stars can form concurrently. Slow prograde stars should form the same way, and these stars should simply extend the extreme $V_\phi$ tail to the other side of the $V_\phi$ = 0 line. These stars may contribute to \textit{Eos} or the alpha-poor component of the \textit{Splash}. The clumpy phase of the galaxy evolution in these simulations generally lasts for $\lesssim 5$ Gyr (see e.g. \citealt{amarante_2020}, \citealt{fiteni_2021}), after which the production of slow and retrograde stars from this mechanism ceases. As this phase would have ended a number of gigayears ago, it is not relevant to estimate a rate as was done with the earlier mechanisms. Here we estimate the approximate fraction of slow and retrograde in-situ stars in the Galaxy today that could be from an early clumpy phase of evolution. 

\citet{fiteni_2021} use isolated, N-body smooth particle hydrodynamic simulations to investigate the relative importance of scattering by early clumps and interactions with galactic bars in producing retrograde stars. Those authors find that the mass fraction of clump-driven retrograde stars --- given as a function of the mass of retrograde stars at a given time to the total mass of all stars at ten gigayears --- stays approximately constant after the end of the clumpy phase. As there are no gas inflows or mergers in the \citet{fiteni_2021} simulations, this mass fraction represents only the in-situ component and does not represent the fraction of clump-driven retrograde stars to the total stellar mass --- making this estimate a high upper bound. Further, there is no distinction between thin and thick disk stars in their analysis. In their simulations, the final clump-driven retrograde (in-situ) mass fraction is $\sim 7\%$. 

Inspection of Figure 4 in \citet{fiteni_2021} indicates that for the barred, clumpy model, there is a clearly-peaked distribution of slow and retrograde stars around $V_\phi \sim 0$ km~s$^{-1}$ while there is not a distinct peak around $V_\phi \sim 0$ km~s$^{-1}$ in the clumpy, unbarred galaxy. Instead, this galaxy has a more continuous distribution of velocities, including those both above and below $V_\phi \sim 0$ km~s$^{-1}$. This figure suggests that there should be at least as many slow prograde stars as retrograde stars, though the abundance properties of these kinematical outliers are unknown. The radial distribution of the clump-driven retrograde stars is peaked at small R ($<5$ kpc, see e.g. Figures 4 and 5 in \citealt{fiteni_2021}). As seen in their Figure 6, the clumpy disk mechanism can scatter retrograde stars to large heights (z $>$ 5~kpc), and the same is presumably true for slow stars. 

Those authors determine that within the Solar cylinder (defined to be $7.5 < \rm{R}_G < 8.5$ kpc and $0 < |\rm{z}| < 2$ kpc), clump-driven retrograde stars in their simulations contribute less than $0.05\%$ of the total in-situ mass. Even if considering a larger range of radii, heights above the plane, and velocity values, the mass fraction of in-situ, clump-driven slow and retrograde velocities is still likely less than $1\%$. Given that the Milky Way has accreted mass throughout its lifetime (and that it is unknown if it had a phase of clumpy star formation) the true percentage should be even lower. 

\eject
\subsubsection{Possible Observational Signatures}
Stars that were scattered by clumps early on in the Galaxy's evolution should be old. The clumps in the \cite{fiteni_2021} simulations exist for $\lesssim 5$ Gyr, which suggests that clump-driven slow and retrograde stars observed today would likely be $\gtrsim 8$~Gyr old.

\subsubsection{Identification of Candidate Stars}
We consider this mechanism to be plausible for only old stars and adopt a generous cut of $> 6$ Gyr to account \texttt{astroNN}'s bias towards younger ages. We rule out a clump-driven origin for stars with ages $\le 6$ Gyr, and do not consider any stars to be consistent with this mechanism.  

\subsection{Interactions with the Galactic Bar}
The Milky Way is a barred galaxy (see e.g. \citealt{blitz_1991}, \citealt{weinberg_1991}), and dynamical interactions with the bar and its associated resonances can alter stellar orbits. Indeed, simulations predict that as the bar and galaxy co-evolve, some number of stars will be instantaneously retrograde at any given point in time due to interactions with the galactic bar (see e.g. \citealt{fiteni_2021}). By extension, some number of stars will have their $V_\phi$ velocities changed to low, barely prograde values. Observed slow or retrograde velocities could possibly be due to torques by the bar or resonant trapping.

\cite{fiteni_2021} find that over ten gigayears of isolated evolution, the stellar bar can cause up to $\sim 14\%$ of the total mass of stars --- again stated as a function of the mass of retrograde stars in a snapshot to the total mass of all in-situ stars at ten gigayears --- to become retrograde. The mass fraction of retrograde stars appears to increase so long as there is a bar within the galaxy. Without more detailed analysis of the relationship between retrograde stars and bar properties over a wider range of parameters --- the subject of our future work --- it is difficult to estimate the in-situ or total mass fraction of bar-induced slow and retrograde stars in the Milky Way. However, from the analysis presented in \cite{fiteni_2021}, we can make predictions about where such stars would be in the Galaxy. 

Those authors find that bar-driven retrograde stars stay generally within the radial extent of the bar, even after dissolution of the bar. The radial and vertical extents are smaller for the bar-driven retrograde stars compared to the clump-driven retrograde stars. Bar-induced, slow prograde stars are likely also similarly confined to the inner galaxy within the radial extent of the bar, though they are not explicitly investigated in \cite{fiteni_2021}. In their Figure 4, there is no clear second velocity peak around $V_\phi \sim 0$ km~s$^{-1}$ for the bar-driven retrograde stars, and at radii greater than $\sim 2-3$ kpc, the velocity distribution is distinctly above the $V_\phi \sim 0$ km~s$^{-1}$ line.

\subsubsection{Possible Observational Signatures}
The stars that become slow prograde or retrograde due to interactions with the bar should generally stay within the extent of the bar, though they can have some small spread above the plane ($\sim 1$ kpc, see Figure 6 in \citealt{fiteni_2021}). The present-day radii of these stars may not be significantly different from their birth radii, though the bar tends to cause stars to move to smaller radii in the inner galaxy (e.g. \citealt{filion_2023}). Stars in the bar also generally follow the inner disk abundance trends discussed above (e.g. \citealt{horta_2024}).

\subsubsection{Identification of Candidate Stars}
The review presented in \cite{bland_hawthorn_2016} gives a Milky Way bar length estimate of $\sim 5$~kpc, though more recent estimates have suggested that the bar may be as short as $\sim 3.5$~kpc \citep{lucey_2023}. There is only one star in the sample with $\rm{R} < 5$, though this star is relatively iron- and alpha-poor ([Fe/H] $\sim -0.5$) and is thus unlikely to be associated with the bar. Even if considering a more generous range of Galactocentric radii, there are only five other stars with R $< 6$ kpc. Of these five stars, two have approximately Solar or super-Solar iron abundances. These two stars could plausibly be related to the bar, though it is not particularly likely. We consider this mechanism to be plausible for these stars, and unlikely for all other stars.

\subsection{Satellite Galaxy Interactions}
A passing satellite galaxy could cause a disk star to become slowly prograde or retrograde if the conditions of the interaction are sufficient to induce a large enough change in the azimuthal component of the velocity of the star ($\Delta \rm{V}_{\phi}$, for additional discussion on the impact of satellite passage on disk orbits see e.g. \citealt{quillen_2009}, \citealt{bird_2012}, and \citealt{carr_2022}). We again assume a nominal star to have $V_\phi \sim -230$ km~s$^{-1}$, $V_R = V_z = 0$ km~s$^{-1}$, and again a rate estimate is not relevant for the temporally discreet event of a satellite passage. The amplitude of the resulting change in velocity depends on the mass of the satellite, the relative velocity of the encounter, and the distance to the satellite. For this scenario, we are particularly interested in close pericenters or disk crossings, and the Sagittarius Dwarf galaxy (Sgr) is the most plausible candidate for producing slow and retrograde stars near the Sun. Sgr is one of the Milky Way's more massive satellite companions, and it likely crossed through the disk mid-plane in the Galactic anti-center direction at least once within the past $\sim 1$ Gyr (see e.g. \citealt{vasiliev_2020} and references therein). Here we investigate the maximum possible $\Delta V_\phi$ due to a disk mid-plane passage of Sgr to determine if this mechanism is feasible. We detail the geometry of the interaction and derive the expression for the change in $V_\phi$ in Appendix \ref{sec:appendix_sgr}.

The largest $\Delta \rm{V}_\phi$ will occur at the highest Sgr mass and slowest disk crossing velocity. The present-day total mass of Sgr is inferred to be $\sim 4 \times 10^8 \rm{M}_\odot$ \citep{vasiliev_2020}. Given that Sgr is tidally stripping, the satellite may have been more massive at the most recent passage, and we generously estimate the mass at the mid-plane crossing to be $1 \times 10^{9} \rm{M}_\odot$. We adopt a Galactocentric radial distance for Sgr ($R_{sgr}$) at the most recent disk mid-plane crossing of 15 kpc, and adopt a velocity ($V_{sgr}$) estimate of 150~km~s$^{-1}$ in only the z-direction (see e.g. the range of recent pericentric passages and integrated \lq present day' kinematic properties presented in \citealt{bennett_2022}). We compute the $\Delta V_\phi$ values for disk stars at z $= 0$ kpc on a grid of $\phi$ and Galactocentric radii values, from $-\pi$ to $\pi$ spaced by 0.5 degrees and from 0.1 to 25 kpc, spaced by 0.1 kpc, respectively. We avoid the exact radius corresponding to where the disk crossing occurs.

The maximum $\Delta \rm{V}_\phi$ for a disk star is $|\Delta \rm{V}_\phi|_{max} \sim 260$ km~s$^{-1}$. Note, however, that this represents a rather extreme set of parameter estimates --- we have simply established that it is within the realm of possibility that an encounter with a Sgr-mass galaxy could cause sufficiently large changes in velocity, not whether it is likely to have happened.\footnote{We note that many simulations find a present-day mass of Sgr that is larger than the mass inferred from observations, which would increase $|\Delta \rm{V}_\phi|$. For example, the simulation from \cite{laporte_2018} that is employed in \cite{carr_2022} has a present-day Sgr mass of $\sim 6 \times 10^9 \rm{M}_\odot$.} We stress that this estimation is oversimplified but helpful for intuition-building. Finally, we note that the passage of Sgr would likely alter the vertical extent of Milky Way disk star orbits. We do not estimate this vertical change here.

\begin{figure*}
    \includegraphics[width=\textwidth]{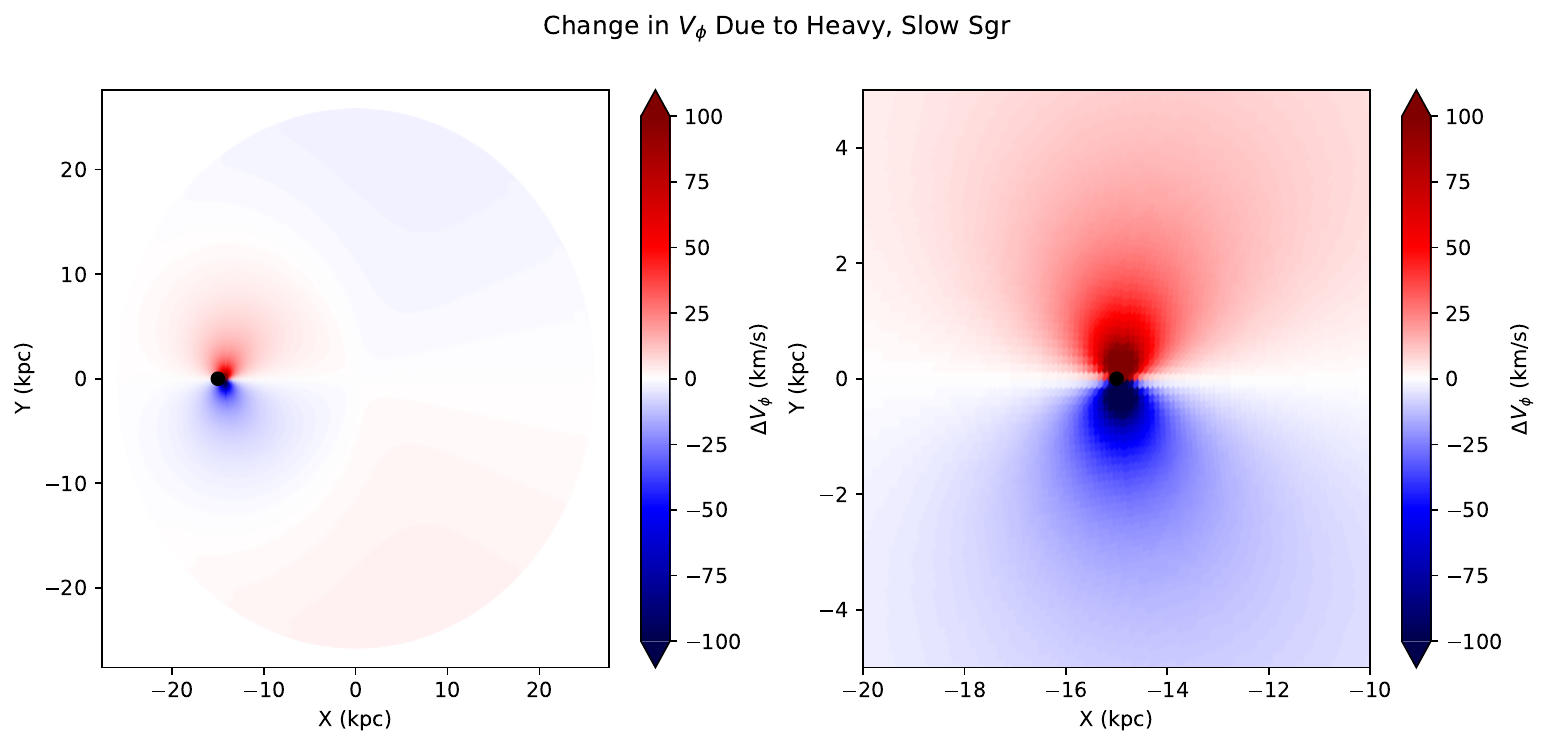}
    \caption{The amplitude and sign of $\Delta \rm{V}_\phi$ computed on a grid in radius and $\phi$ for a set of parameters that can provide high amplitude $\Delta \rm{V}_\phi$ ($M_{sgr} = 1\times10^9 \rm{M}_\odot$, $R_{sgr} = 15$ kpc, and $V_{sgr} = 150$ km~s$^{-1}$). The plot at left shows the full disk, while the plot at right shows a zoom-in on the region closest to the location of the satellite passage (shown in both panels with a black circle), where the gridding can be seen more clearly. In both panels, the maximum (minimum) of the color bar has been set to 100 (-100) km~s$^{-1}$ and all values above (below) these values are colored as maximum (minimum) value.} 
    \label{fig:delta_vphi}
\end{figure*}

As seen in Figure \ref{fig:delta_vphi}, the largest amplitude $\Delta \rm{V}_\phi$ will be closest to where the disk crossing occurs. Even in the idealized case (i.e. maximal $M_{sgr}$ and minimal $V_{sgr}$ and $R_{sgr}$), only a small area of the disk would have sufficiently large $\Delta \rm{V}_\phi$ values to produce retrograde stars. For example, if we adopt the values given above, the maximal $|\Delta \rm{V}_\phi|$ for a star at $R_{star} = 14.8$ kpc, $|\Delta \rm{V}_\phi|$ is $\sim 140$ km~s$^{-1}$, while a star at $R_{star} = 14.5$ kpc the maximum is $\sim 60$ km~s$^{-1}$. The density distribution of stars in the disk declines as a function of radius, such that we would expect very few ($< 1\%$ of disk stars) to be within the area with sufficiently large $|\Delta \rm{V}_\phi|$ values.

\subsubsection{Possible Observational Signatures}

Given the argument above, shortly after pericentric passage we would only expect satellite-induced slow prograde and/or retrograde stars in the outer disk, concentrated within a relatively small area. There would also be stars nearby with faster-than-normal $V_\phi$ due to additive $\Delta \rm{V}_\phi$ values. Phase mixing would erase the spatially coherent signal (see Fig. \ref{fig:delta_vphi}) over time, but the perturbed velocities would remain. As discussed in \cite{carr_2022}, the satellite torque would likely heat orbits and also cause stars to move radially. Those authors note that this radial motion could cause an azimuth-angle dependent pattern in the metallicity of stars at a given Galactocentric radius which would then phase mix away.

\subsubsection{Identification of Candidate Stars}
As Sgr likely passed through the outer disk, any slow and retrograde stars produced from this encounter would likely have sub-Solar metallicities given the radial metallicity gradient. We can thus reject the possibility that this mechanism is responsible for the slow and retrograde metal-rich stars. The combination of a sufficiently heavy, slow Sgr passing through radii within the extent of the disk may not be particularly likely. However, given the uncertainties on the orbit of Sgr, we consider this mechanism to be plausible, but not likely (i.e. not consistent), for the stars with iron abundances below Solar. We consider this mechanism to be unlikely for all other stars.


\section{Discussion}\label{sec:discussion}
\begin{figure*}
   \includegraphics[width=\textwidth]{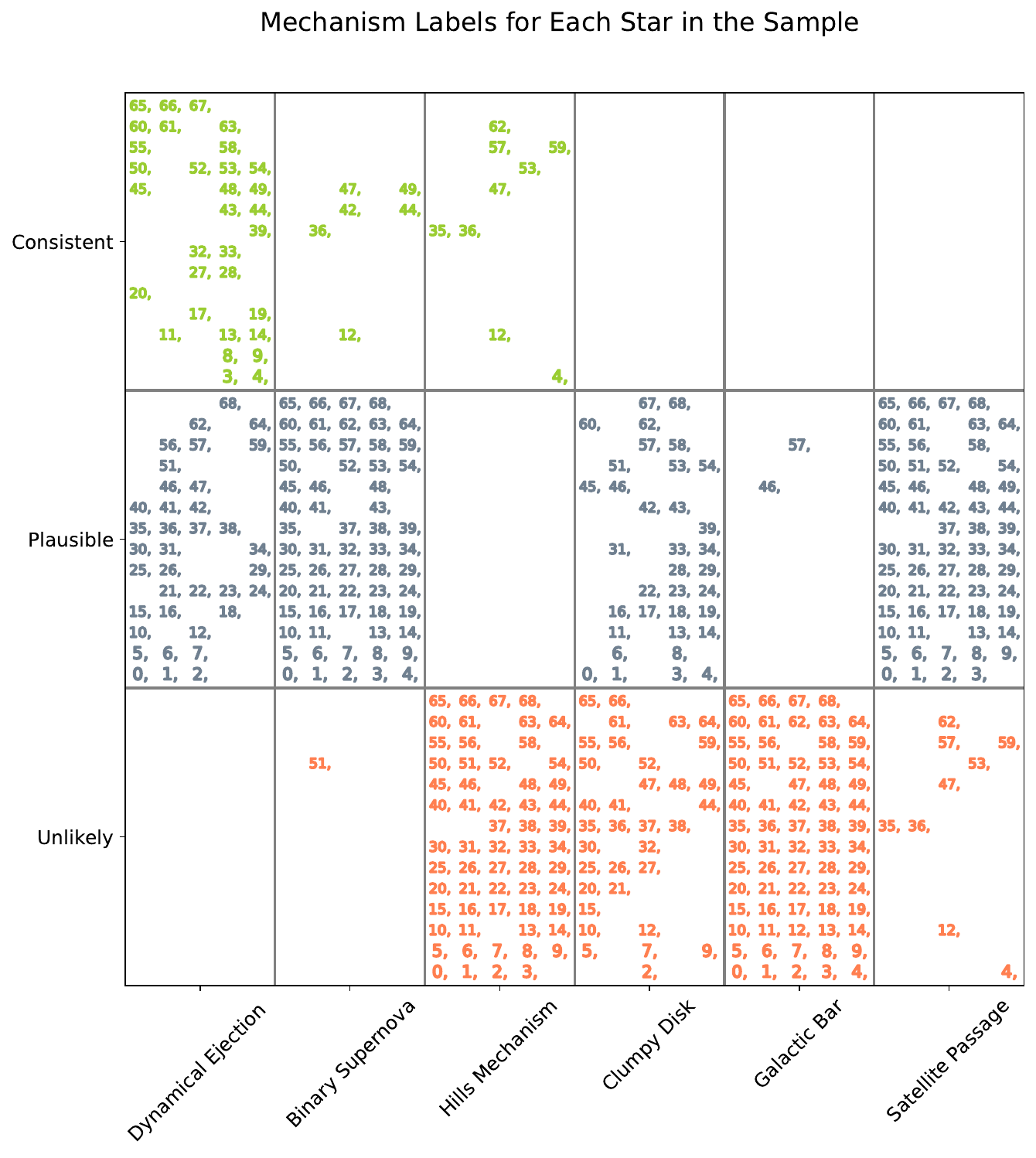}
    \caption{The final labels for each mechanism for each star, where each number corresponds to a unique APOGEE ID. The correspondence between number and APOGEE ID is given in Table \ref{tab:key}.}
    \label{fig:results}
\end{figure*}

\begin{figure}
   \includegraphics[width=.475\textwidth]{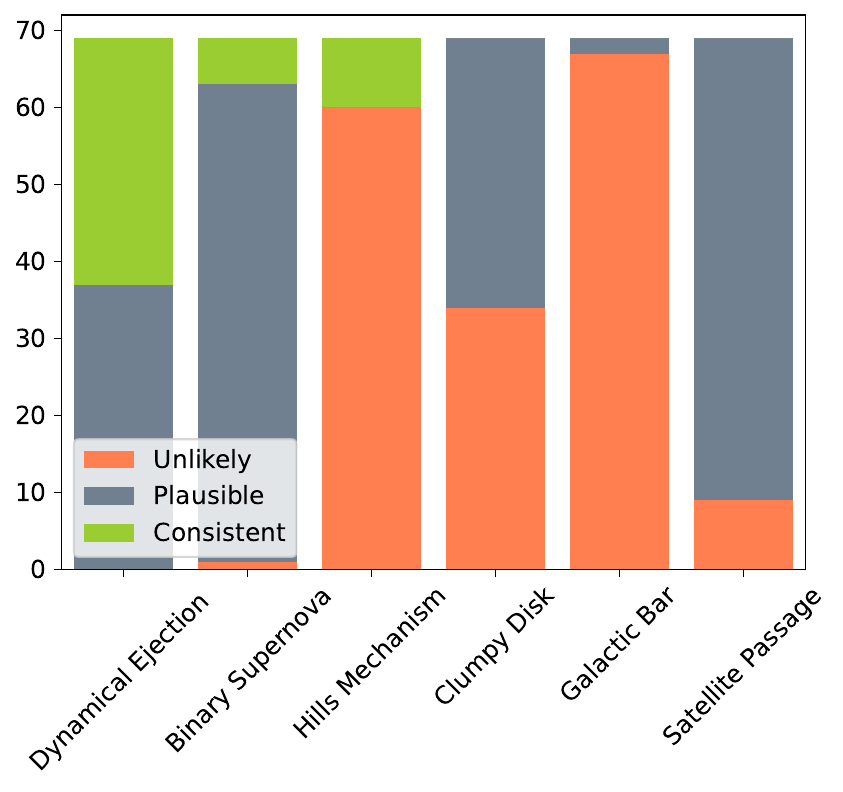}
    \caption{A stacked bar chart showing the number of unlikely (coral), plausible (grey), and consistent (green) stars for each mechanism discussed in the text.}
    \label{fig:barchart}
\end{figure}

From the above mechanisms for which an ejection rate estimate is possible and meaningful, the Hills mechanism is capable of producing the most slow or retrograde stars, followed by binary supernova ejection, then dynamical cluster ejection. The clumpy disk and satellite passage mechanisms each likely produce fewer slow and retrograde stars than the Galactic bar, if the bar is not very young. We summarize the conclusions of the previous section in Table \ref{tab:results}, Figure \ref{fig:results}, and Figure \ref{fig:barchart} (see also supplementary Table \ref{tab:key} for APOGEE IDs for each star). Here, a mechanism is labeled \lq unlikely' for a given star if the star has properties that are counter to that mechanism's signatures --- for example, metal-poor stars are considered unlikely Hills mechanism candidates. A mechanism is considered \lq plausible' if a given star displays no clear, strong signatures of the mechanism but the mechanism remains plausible. Finally, a mechanism is \lq consistent' if a star has properties that match the signatures given above. Some mechanisms lack strong signatures, such as the early clumpy disk and the Sgr crossing scenarios, which means that there are no \lq consistent' candidates. 

\begin{table*}[]
\resizebox{\textwidth}{!}{%
\begin{tabular}{|l|l|l|l|}
\hline
                   & \multicolumn{1}{c|}{Consistent}                      & \multicolumn{1}{c|}{Plausible} & \multicolumn{1}{c|}{Unlikely} \\ \hline
Dynamical Ejection & Abundances similar to existing & Properties are non-constraining               & No excluding properties                          \\ 
 &  globular clusters &    (possible for all stars)                                    &                           \\ \hline
Binary Supernova   & Low {[}C/N{]} and/or high    & No binarity      & Binarity                      \\ 
   &  rotational velocity    &      &                  \\ \hline
Hills Mechanism    & Super-solar metallicity, eccentric orbit                          & N/A                          & Low metallicity               \\ \hline
Clumpy Disk        & No strong signatures                             & Old age estimate              & Young age estimate            \\ \hline
Galactic Bar &
  Located within the bar length and near &
  Located slightly outside of the bar and  &
  All stars outside of bar region, low \\ 
 &
  Solar or super-Solar metallicity &
   near Solar or super-Solar metallicity &
   metallicity stars in or near bar region \\ \hline
Satellite Passage &
  No strong signatures &
  Sub-Solar metallicity &
  Solar or super-Solar metallicity \\ \hline
\end{tabular}%
}
\caption{Summary of the the properties determining whether a star was considered consistent, plausible, and unlikely for each mechanism. Note that while we provide the requisite properties, there are no consistent stars for the Galactic bar mechanism.}
\label{tab:results}
\end{table*}

While at least one mechanism is at a minimum plausible for each star, the vast majority of the slow and retrograde stars were likely not affected by the Galactic bar or the Hills mechanisms. It is difficult to rule out the binary supernova or cluster ejection scenario for most stars, though there exists a subset of stars with abundance indicators that are consistent with possible expected signatures of these mechanisms. The scenario with the highest number of consistent candidates is the dynamical ejection scenario, this label is driven by the abundance similarity between stars in the sample and globular clusters in APOGEE. Interestingly, low [C/N] abundances could be indicative of either binary interactions or globular cluster ejection, given the range of [C/N] seen in NGC6388 (Figure \ref{fig:simchem}, remembering that low carbon and high nitrogen abundances are a signature of second generation globular cluster stars, e.g. \citealt{Bastian2018}). Finally, we note that a given star could have been effected by more than one of the mechanisms discussed above (and/or two Sgr passages), though the probability of two rare events happening to the same star are exceedingly low.

Given the relatively low predicted production rate, it is perhaps surprising that the dynamical ejection mechanism has the largest number of consistent candidate stars. However, these rate estimates are meant only to provide an order-of-magnitude sense of the landscape, and in this regard the rates are not dramatically different --- they vary only by a factor of $\sim $2-4. A number of uncertainties can increase or decrease the production rate of slow and retrograde stars from dynamical ejection. For example, the potential parent globular clusters may have been on slower, more favorable orbits at the time of ejection. The simulations presented in \citet{schoettler_2022} suggest that the fractal nature of star formation can influence the rate and possibly velocity distribution of ejected stars, which suggests that including more substructure in the early stages of cluster evolution could alter the number of sufficiently fast ejections. The binary supernovae and Hills mechanism rate estimates above are also impacted by uncertainties: for example, the binary ejection mechanism is sensitive to uncertain binary evolution physics \citep{evans_2020}, and the orbits of stars post-ejection from the Hills mechanism is sensitive to the Galactic potential. As noted above, recent analyses suggest that the Hills mechanism ejection rates may be lower than that assumed here. Through further analyses linking both slow/retrograde stars and fast/hypervelocity stars to their origins, it may be possible to reduce these uncertainties and gain a better understanding of the different production processes.

In this analysis, stars are tied to mechanisms based on their various properties. In a more broad-brush sense, we can also consider mechanisms as a function of metallicity. Lower metallicity stars, for example, are more likely to be linked to the satellite passage mechanism due to the radial metallicity gradient in the disk. Old stars are more likely to be metal-poor, and old stars in this sample are associated with the early clumpy disk scenario (note, however, that stars with similar \texttt{astro\_nn} age estimates span a wide range in abundances, as seen in Figure \ref{fig:abundances_reliable}). Fast binary supernovae ejections may also be more likely at lower metallicity (e.g. \citealt{evans_2020}). Similarly, a metallicity dependence in dynamical ejections could possibly arise due to metallicity-dependent close binary fractions \citep[e.g.,][]{moe_2019, wyse_2020}. Higher metallicity stars are more likely to be tied to the Hills mechanism and/or the Galactic bar mechanism.

We can also consider the ejection mechanisms as a function of time. Here we again stress that \texttt{astro\_nn} ages are based on spectra, and the accuracy of any given age estimate cannot be guaranteed, especially for stars with atypical abundances.  It seems likely that dynamical ejection rates would be higher in past eras of  higher star formation, under the assumption that higher star formation tends to mean more clusters. High-mass primaries and binary supernovae are also likely more common in more active periods of star formation. The rate from the Hills mechanism may not be constant in time, but it is unclear if this rate should be higher or lower in the past. As discussed above, the clumpy disk scenario is only possible during early stages of Galactic formation. The number of retrograde stars due to the Galactic bar should increase with time \citep{fiteni_2021}, though the exact age of the Galactic bar is not known. Finally, the Sgr crossing is discrete in time. 

Thus far, all of the mechanisms introduced implicitly assume that a given star was born in the disk and subsequently had its orbit altered. Under this assumption, the sample presented here would represent the extreme kinematical tail of the thin disk. However, it is possible that some of the stars may actually be halo stars with thin-disk-like abundances. These stars could possibly be the low-alpha, extreme tail of either an accreted population or the in-situ halo. We explore each of these possibilities in the following subsections.

\subsection{An Ex-Situ Halo Origin}
The thin disk is chemically evolved, and any accreted population with overlap in the [Mg/H]--[Mg/Fe] and [Al/Fe]--[Mg/Mn] planes could only come from the more massive satellites, as shown in \cite{hasselquist_2021} and \cite{fernandes_2023}. The abundance distributions presented in those analyses suggest that only the tails of the GSE and Large Magellanic Cloud (LMC) distributions could possibly overlap with the low [Al/Fe] end of the thin disk distribution. However, we posit that it is unlikely that the stars in the slow and retrograde sample came from either of these galaxies.

As shown in (e.g.) \cite{carrillo_2024}, the exact choice of how candidate GSE stars are selected impacts the resulting abundance distribution. Selections only in angular momentum or eccentricity will likely select stars from the slow and retrograde sample presented here, and as illustrated above, these properties alone do not necessarily indicate an ex-situ in origin. For the most restrictive action-based selections, there are likely few to no candidate GSE stars that overlap with the thin disk regions of the [Mg/H]--[Mg/Fe] and [Al/Fe]--[Mg/Mn] planes. Indeed, the selection of GSE stars in \cite{carrillo_2024} that is most similar to that employed in \cite{fernandes_2023} does include some stars in the thin disk region of the [Al/Fe]--[Mg/Mn] plane, while a stringent selection in action space, for example, removes all or nearly all stars from this region of abundance space. The ambiguity of GSE definitions and properties makes it difficult to fully reject the possibility that some of the stars in the slow and retrograde sample were born in the GSE, though the chemically evolved nature of any such stars is perhaps surprising for a dwarf galaxy that presumably stopped forming stars $\sim 10+$ Gyr ago.

LMC heritage for stars in the slow and retrograde sample is also unlikely, though not impossible. Unlike the GSE, the LMC is still largely intact and is likely on its first infall (e.g. \citealt{nitya_2006}). The LMC is tens of kiloparsecs from the Sun, as is any possible stellar stream (e.g. \citealt{petersen_2022}), while as seen in Figures \ref{fig:spatial_0_to_4} through \ref{fig:spatial_7_plus}, the stars in the slow and retrograde stars sample are all relatively nearby. As such, the only possible way for an LMC star to be in the slow and retrograde sample would be if that star were ejected from the LMC via a number of the mechanisms described above. As shown in e.g. \cite{evans_2021}, \cite{boubert_2017}, and \cite{han_2025}, the mechanisms outlined here for producing high-velocity stars can eject stars out of the LMC, though it is rare for ejected stars to end up near the Sun. For an LMC-ejected star to be in the sample, it must be at the more chemically evolved end of the abundance distribution of the LMC, have an ejection speed above the LMC escape velocity, and an ejection time that was long enough ago that the star could travel $\sim 45+$ kiloparsecs to its present location near the Sun. The combination of these requirements make an LMC origin relatively less likely than the other mechanisms described here.

\subsection{An In-Situ Halo Origin}
The possibility of an in-situ halo origin for some stars in the sample is very closely linked to the origins of the \textit{Eos} and \textit{Splash} populations discussed in Section \ref{section:substructure_comp}, as well as the early clumpy disk and Sgr passage scenarios. It is likely that the in-situ halo is comprised largely or entirely of heated proto (thick) disk material \citep[][]{bonaca_2017, DiMatteo2019}, which is also a proposed origin of the \textit{Splash} \citep{belokurov_2020}. An early merger is usually the proposed mechanism for this heating. The thin disk would then subsequently form stars on cooler orbits, possibly from gas deposited by this merger. The other proposed origin for the \textit{Splash} is the clumpy disk mechanism described in Section \ref{sec:mechanisms}. 

We have somewhat artificially separated stars in the sample resulting from either of these two possible in-situ halo/\textit{Splash} origins into elemental abundance halo outliers or kinematical thin-disk outliers, respectively, and discussed the latter in Section \ref{sec:mechanisms}. The logic of this decision is as follows: in the merger-heated proto-disk scenario, the stars in the slow and retrograde sample belong to the population of proto-disk stars that were heated to become halo stars, but they are abundance outliers from the rest of that population. These abundance outliers could even be a sign that some amount of the (abundance-based definition of the) thin disk had already formed by the time of the merger. The clumpy disk, however, produces thin and thick disk stars concurrently, and the slow and retrograde stars are kinematic outliers of the thin disk. 

The distinction between extreme halo and extreme disk stars is most difficult to make at the older, lower metallicity end, which trace a time when the Milky Way was in a more turbulent stage of its development. It is of course also possible that some slow and retrograde stars could have formed during the proposed merger from the gas that went on to produce the thin disk (see also the scenario presented in \citealt{renaud_2021}), which further muddies the distinction between halo and disk stars here. The slow and retrograde stars with thin-disk abundances resulting from any of these halo scenarios would have the same labels as those in the clumpy disk column of Figure \ref{fig:results} and \ref{fig:barchart}, which were decided based exclusively on estimated (old) age. More precise age estimates could help to disentangle these possibilities.

\subsection{Implications}
The stars in this sample present an opportunity to challenge common assumptions. For example, the first three mechanisms discussed in this analysis are generally associated with producing runaway or fast stars. High velocities are an easily identifiable signature of these mechanisms, and fast stars stand out from the nominal background population. However, if the goal is to understand these processes, more complete samples can be made by searching for both fast stars and slow/retrograde stars. In both cases, the ejected stars are kinematically distinct, but the low-to-negative $V_\phi$ end is often neglected. Focusing on the slow or retrograde end allowed us to identify stars that may have been ejected from clusters, supernovae, or the Hills mechanism. 

Many of the mechanisms discussed in this analysis also present alternative avenues for moving stars far from their birth radii, which can impact the tails of local abundance distributions. In particular, dynamical ejection, binary supernovae ejection, and the Hills mechanism are often not considered in discussions of stars with abundances that differ from their similarly-aged peers. The arguments about high ejection velocity (or change in velocity) and direction made throughout this analysis need not apply in this context, increasing the potential impact of these mechanisms on abundance distributions. 

\section{Conclusions}\label{sec:conclusion}

In this work, we have identified and analyzed a sample of $\sim 70$ stars with low-alpha, thin-disk-like abundances and slow or retrograde $V_\phi$ velocities. The stars that exist in this extreme are informative --- they challenge our assumptions and allow us to investigate rare processes. For the majority of the analysis, we assume that the stars in this sample were born with thin-disk-like kinematics and had their orbits altered. We consider six possible mechanisms for altering stellar orbits and determine possible observational signatures of each mechanism. We then consider the possibility that some stars are instead truly halo stars with atypical abundances. We present our results here:

\begin{enumerate}
    \item \textbf{The dynamical ejection scenario}: slow and retrograde stars are ejected via multi-body interactions in dense stellar clusters. We link stars in the sample to possible parent globular clusters with similar abundances, but note that any star could have been ejected by a cluster that has since dissolved. Based on the elemental similarity between our sample and clusters with thin disk chemistry, we find this mechanism to have the largest number of \lq consistent' candidate stars.
    \item \textbf{Binary supernovae mechanism}: slow and retrograde stars are produced when their more massive binary companion goes supernova. This mechanism could induce rotation and/or alter the surface [C/N] abundance of the ejected star, though it is also possible that there could be no lasting observable signature. We thus consider this mechanism to be consistent with stars with high rotation velocities and/or low [C/N] ratios, and plausible for the other single stars in the sample. This mechanism is unlikely for the probable binary star in the sample.
    \item \textbf{Hills mechanism}: slow and retrograde stars are ejected after a stellar binary is disrupted by the central supermassive black hole. The main signature of this mechanism for stars in this sample is high metallicity, as Hills-ejected stars likely formed in the central regions of the Galaxy, and eccentric orbits. We find a number of consistent candidate stars for this mechanism. These candidates could be confirmed with a detailed orbit integration analysis that incorporates an evolving bar.
    \item \textbf{Early clumpy phase of the disk}: slow and retrograde stars are formed from scattering off of star forming clumps. Early massive star forming clumps have been proposed as a potential avenue for creating the thin and thick disks concurrently, as well as a heated \textit{Splash}-like population. These stars may contribute to the \textit{Eos} or alpha-poor component of the \textit{Splash} substructure. While it is unknown if the Milky Way had a clumpy phase, any stars that would have been effected by the clumps would have formed early on. As such, only the old stars in the sample could plausibly have had their orbits altered by this mechanism. There are no clear, strong signatures of this mechanism.
    \item \textbf{Galactic bar interactions}: slow and retrograde stars are produced through dynamical interactions with the Galactic bar. Simulations suggest that stars with retrograde orbits are confined within the bar, and we argue that stars in the bar should generally be Solar or super-Solar. We apply the same criteria for the slow prograde stars. There are only two stars that could plausibly be associated with the bar, though their Galactocentric radii place them nominally outside of the bar. 
    \item \textbf{Disk crossing or pericentric passage of Sgr}: slow and retrograde stars could be caused by the passage of Sgr through the Galactic plane. Under the assumption of the impulse approximation, we find that there are combinations of parameters for Sgr and its orbit that can produce slow and retrograde stars. Any slow and retrograde stars resulting from this scenario would likely have sub-Solar iron abundances that reflect their origin in the outer disk, though there is no other clear signature. However, the number of stars that would be effected is low and the parameter combinations may not be particularly likely.
    \item Finally, we consider the possibility that some slow and retrograde stars are truly halo stars. These stars could be ex-situ or in-situ, with in-situ being more likely. These stars would likely be old and are thus the same set of stars for which the clumpy disk scenario is plausible.
\end{enumerate}

It is unlikely that a single mechanism is responsible for the altering the orbits of each of the stars in this sample. Instead, it is likely a combination of a subset or all of the scenarios given here. The next few years will provide more data within which to search for outlier stars. For example, the Sloan Digital Sky Survey Five (SDSS-V, \citealt{sdssv}) and the Multi-Object Optical and Near-IR Spectrograph
(MOONS, \citealt{moons_2,moons}) will be able to provide more data for stars towards the Galactic center, while Prime Focus Spectrograph (PFS, \citealt{pfs}) will observe stars in the Galactic outskirts. These surveys, along with further data releases from \textit{Gaia} and future astrometric surveys, will help us better understand the Milky Way by enabling the discovery of more oddballs with surprising abundances and kinematics.

\appendix

\section{$\Delta \rm{V}_\phi$ Impulse Approximation Calculation}\label{sec:appendix_sgr}
We estimate the amplitude of $\Delta \rm{V}_{\phi}$ for disk stars in a Milky Way-like galaxy during a close passage of Sgr-like galaxy. The computation and coordinate orientation largely follows that of \cite{gandhi_2022} and \cite{carr_2022}, which we briefly summarize here. We follow \cite{carr_2022} and adopt the impulse approximation, which should be reasonably accurate for the intended purpose of estimating the magnitude of the change in velocity. As those authors note, the impulse approximation allows us to ignore the motions of stars within the disk and thus set the Galactocentric (cylindrical) radii of stars and Sgr to their values at pericenter or mid-plane disk crossing. We assume Sgr moves only in the z-direction.

We first assume that Sgr is located along the x-axis at $R_{sgr}$ away from the origin, such that its (x, y) position is ($R_{sgr}$, 0). We then assume that Sgr is moving along the $z$-axis with a velocity of $V_{sgr}$, such that its $z$-position at any given time is $V_{sgr}t$. We assert that all stars are in the plane of the Galaxy, such that the position vector between a given star and the Galactic center is $\vec{r}_\star = [R_\star \cos(\phi) \ \hat{x}, \ R_\star \sin(\phi) \ \hat{y}, \ 0 \ \hat{z}]$. Here $R_\star$ can be any value between 0 kpc and $\sim 25$ kpc, corresponding to the edge of the disk. $\vec{r}_{sgr}$ is [$R_{sgr} \ \hat{x}, \ 0 \ \hat{y}, \ \rm{V}_{sgr} t \ \hat{z}$]. The position vector between the star and Sgr is then $\vec{r} = [(R_{sgr} - R_\star \cos(\phi)) \ \hat{x}, \ - R_\star \sin(\phi) \ \hat{y}, \ V_{sgr} t \ \hat{z}]$, remembering that Sgr is along the x-axis. Equation 2 of \cite{carr_2022} describes the tidal acceleration of a given star in the disk --- accounting for the individual star and the acceleration of the Galactic center --- and we insert these vectors into their equation. We reorganize slightly such that we begin with: 
\begin{equation}\label{equation:acc}
a_t(t) = G M_{sgr} \left(-\frac{\vec{r}}{|r|^3}+\frac{\vec{r}_{sgr}}{|r_{sgr}|^3}\right) .
\end{equation}
To obtain the impulsive change in the velocity due to the passage of Sgr, we integrate Equation \ref{equation:acc} over time: $\int_{-\infty}^{+\infty} a_t(t) dt$. Integrating the x- and y- components gives $\Delta \rm{v}_x$, $\Delta \rm{v}_y$, as shown in Equations \ref{equation:vx} and \ref{equation:vy}. We then transform the resulting $\Delta \rm{v}_x$, $\Delta \rm{v}_y$ into $\Delta \rm{V}_\phi$ and investigate its maximum amplitude for a range of physically plausible choices of $M_{sgr}$, $V_{sgr}$, and $R_{sgr}$. $\Delta \rm{V}_\phi$ will be highest closest to $\rm{R}_{sgr}$, and the maximum $\Delta \rm{V}_\phi$ value will come from the combination with the highest $M_{sgr}$ value and smallest $\rm{V}_{sgr}$ and $R_{sgr}$ values.

\begin{equation}\label{equation:vx}
\frac{\Delta \rm{v}_x}{G M_{sgr}} = \frac{2 R_{\star} (R_{\star} - R_{sgr} \cos(\phi))}{R_{sgr} V_{sgr} (R_{\star}^2 + R_{sgr}^2 - 2 R_{\star} R_{sgr}\cos(\phi))}
\end{equation}

\begin{equation}\label{equation:vy}
\frac{\Delta \rm{v}_y}{G M_{sgr}} = \frac{ 2 R_{\star} \sin(\phi)}{V_{sgr} (R_{\star}^2 + R_{sgr}^2 - 2 R_{\star} R_{sgr} \cos(\phi))}
\end{equation}

\begin{center}
\begin{longtable}{ |c|c|c|c| }
\hline
Number  &   APOGEE ID  &  Number  &   APOGEE ID \\
\hline
0  &  2M00014230+1702419  &  35  &  2M15023086+4212111 \\
1  &  2M00303867+1549501  &  36  &  2M15023130+5112297 \\
2  &  2M01231405-0102329  &  37  &  2M15050196-4430509 \\
3  &  2M01244485+1453267  &  38  &  2M15070607+2222028 \\
4  &  2M01590930+3802046  &  39  &  2M15104604+3158498 \\
5  &  2M04022073-7023289  &  40  &  2M15233167+0919257 \\
6  &  2M04215815-6157227  &  41  &  2M15334993+0934139 \\
7  &  2M06234629-7005024  &  42  &  2M15490280+2650095 \\
8  &  2M07472456+5311408  &  43  &  2M15513506+2805029 \\
9  &  2M08372116-8537442  &  44  &  2M15544945+2915314 \\
10  &  2M09191140+3346373  &  45  &  2M15572935+2608034 \\
11  &  2M09460164+3322570  &  46  &  2M16003181-2139519 \\
12  &  2M10391298+2948411  &  47  &  2M16013180+4056403 \\
13  &  2M10475148-4448333  &  48  &  2M16032054+3918321 \\
14  &  2M11291121-8501189  &  49  &  2M16045588+2738536 \\
15  &  2M12325862+3626584  &  50  &  2M16065591-2222130 \\
16  &  2M12414587+4926490  &  51  &  2M16082687+1809497 \\
17  &  2M13001371+5217200  &  52  &  2M16132140+5253093 \\
18  &  2M13020472+5743438  &  53  &  2M16215605+4725419 \\
19  &  2M13041724-0134399  &  54  &  2M16434878+3726165 \\
20  &  2M13123688+1801404  &  55  &  2M16453831+4802186 \\
21  &  2M13323559-0118149  &  56  &  2M17100538+6323032 \\
22  &  2M13381907+2938588  &  57  &  2M17121198-2438245 \\
23  &  2M13403109-1550124  &  58  &  2M17200667+4207505 \\
24  &  2M13423397+2745255  &  59  &  2M17252566+5837413 \\
25  &  2M13453784-1539189  &  60  &  2M18212748-3903116 \\
26  &  2M13481254-0053486  &  61  &  2M18451564+0557157 \\
27  &  2M14003743+0334382  &  62  &  2M18555040-2808410 \\
28  &  2M14005878+0123010  &  63  &  2M20503798-0045260 \\
29  &  2M14025155+4747385  &  64  &  2M22095666-3445576 \\
30  &  2M14044997+0320333  &  65  &  2M22193340-4820346 \\
31  &  2M14070704+5411394  &  66  &  2M22221939-5008474 \\
32  &  2M14130012+0544007  &  67  &  2M23110573-5259267 \\
33  &  2M14575158+5300140  &  68  &  2M23482381+8501382 \\
34  &  2M15022633+3248296  &  \phn  &  \phn \\
\hline
\caption{The APOGEE IDs corresponding to the numbers given in Figure \ref{fig:results}}
\end{longtable}\label{tab:key}
\end{center}

\begin{acknowledgments}
The authors thank the referee for their helpful suggestions. The authors would like to thank Newlin Weatherford and Fraser Evans for providing both their simulation outputs and constructive commentary. CF thanks the Nearby Universe and Data groups at the CCA for engaging discussion on these topics. CF also thanks Joseph Long for his suggestions on Figure 9, as well as Rachel McClure and Catherine Manea for their thoughts and encouragement. MSP acknowledges support from a UKRI Stephen Hawking Fellowship.  ML is supported by the National Science Foundation under Award No. 2303831. DH was supported by the UKRI Science and Technology Facilities Council under project 101148371 as a Marie Curie Research Fellowship.

Funding for the Sloan Digital Sky Survey IV has been provided by the Alfred P. Sloan Foundation, the U.S. Department of Energy Office of Science, and the Participating Institutions. SDSS acknowledges support and resources from the Center for High-Performance Computing at the University of Utah. The SDSS web site is www.sdss4.org. SDSS is managed by the Astrophysical Research Consortium for the Participating Institutions of the SDSS Collaboration including the Brazilian Participation Group, the Carnegie Institution for Science, Carnegie Mellon University, Center for Astrophysics | Harvard \& Smithsonian (CfA), the Chilean Participation Group, the French Participation Group, Instituto de Astrofísica de Canarias, The Johns Hopkins University, Kavli Institute for the Physics and Mathematics of the Universe (IPMU) / University of Tokyo, the Korean Participation Group, Lawrence Berkeley National Laboratory, Leibniz Institut für Astrophysik Potsdam (AIP), Max-Planck-Institut für Astronomie (MPIA Heidelberg), Max-Planck-Institut für Astrophysik (MPA Garching), Max-Planck-Institut für Extraterrestrische Physik (MPE), National Astronomical Observatories of China, New Mexico State University, New York University, University of Notre Dame, Observatório Nacional / MCTI, The Ohio State University, Pennsylvania State University, Shanghai Astronomical Observatory, United Kingdom Participation Group, Universidad Nacional Autónoma de México, University of Arizona, University of Colorado Boulder, University of Oxford, University of Portsmouth, University of Utah, University of Virginia, University of Washington, University of Wisconsin, Vanderbilt University, and Yale University.

KJD respectfully acknowledges that the University of Arizona is on the land and territories of Indigenous peoples. Today, Arizona is home to 22 federally recognized tribes, with Tucson being home to the O’odham and the Yaqui. 
The University strives to build sustainable relationships with sovereign Native Nations and Indigenous communities through education offerings, partnerships, and community service.
We respect and honor the ancestral caretakers of the land, from time immemorial until now, and into the future.

This work has made use of data from the European Space Agency (ESA) mission
{\it Gaia} (\url{https://www.cosmos.esa.int/gaia}), processed by the {\it Gaia} Data Processing and Analysis Consortium (DPAC, \url{https://www.cosmos.esa.int/web/gaia/dpac/consortium}). Funding for the DPAC has been provided by national institutions, in particular the institutions participating in the {\it Gaia} Multilateral Agreement.
\end{acknowledgments}

%

\vspace{5mm}


\software{astropy \citep{astropy:2013,astropy:2018,astropy:2022}, gala \citep{adrian_price_whelan_2020_4159870,gala}}
\facilities{Du Pont (APOGEE), Sloan (APOGEE), Spitzer, WISE, 2MASS, Gaia}


\bibliography{clean_text}{}
\bibliographystyle{aasjournal}



\end{document}